%% file: main.tex
\documentclass[lettersize,journal]{IEEEtran}
\usepackage{amsmath,amsfonts}
\usepackage{algorithmic}
\usepackage{algorithm}
\usepackage{array}
\usepackage{textcomp}
\usepackage{stfloats}
\usepackage{url}
\usepackage{verbatim}
\usepackage{graphicx}
\usepackage{cite}
\usepackage{graphicx}
\usepackage{balance} 
\usepackage[multiple]{footmisc}
\usepackage{url}
\usepackage{textcomp}
\usepackage{listings}
\usepackage{caption}
\usepackage{subcaption}
\usepackage{float} 
\usepackage[group-separator={,},group-minimum-digits={3}]{siunitx}
\usepackage{enumitem}
\usepackage{amsmath,bm}
\usepackage{eurosym}
\usepackage{multicol}
\usepackage{multirow}
\usepackage{booktabs}
\usepackage{threeparttable}
\usepackage{hyperref}
\usepackage{makecell}
\usepackage[table]{xcolor} 
\usepackage{xcolor}
\usepackage[most]{tcolorbox}
\usepackage{tabularx}

\usepackage{tcolorbox}
\tcbuselibrary{skins}

\makeatletter
\newcommand{\stitle}[1]{\smallskip\noindent\textbf{#1\@addpunct{.}}}

\newcommand\sys{\textsc{Freyja}} 
\newcommand\sysbm{\sys$_{\text{BM}}$}
\newcommand{\revision}[1]{#1}

\tcbset{
    examplebox/.style={
        enhanced,
        colback=red!10!white, 
        boxrule=0pt, 
        fonttitle=\bfseries,
        sharp corners,
        colbacktitle=red!20!white,
        coltitle=red!75!black,
        detach title,
        before upper={\tcbtitle\quad},
    }
}

\newtcolorbox{examplebox}[1]{
    examplebox,
    title=#1
}

\begin{document}

\title{\sys: Efficient Join Discovery in Data Lakes}

\author{Marc Maynou$^1$, Sergi Nadal$^1$, Raquel Panadero$^2$, Javier Flores$^1$, Oscar Romero$^1$, Anna Queralt$^1$\\
$^1$Universitat Politècnica de Catalunya, Barcelona, Spain. $^2$TU Wien, Vienna, Austria\\
\{marc.maynou, sergi.nadal, javier.flores, oscar.romero, anna.queralt\}@upc.edu, raquel.palenzuela@tuwien.ac.at\\

    
}



\maketitle

\begin{abstract}

We study the problem of efficiently computing rankings of joinable attributes in data lakes. Traditional set-overlap measures produce numerous false positives in this scenario, while modern, more accurate Table Representation Learning (TRL) techniques incur prohibitive computational costs. 
In contrast to the state-of-the-art, we adopt a novel notion of join quality tailored to data lakes relying on a metric that combines multiset Jaccard and cardinality proportion. The proposed metric merges the best of both worlds by leveraging syntactic measures while achieving accuracy scores comparable to those of TRL approaches. Generating rankings of joinable pairs is highly scalable at both preparation and query time, since we train a general-purpose predictive model. Predictions are based on data profiles, succinct and efficiently computed representations of dataset characteristics.  
Our experiments show that our system, \sys, matches and improves upon, the results obtained by the state-of-the-art while reducing execution costs by orders of magnitude. 
\end{abstract}


\input{1_introduction}
\input{2_relatedwork}
\input{3_scalable_join_discovery}
\input{4_join_quality_metric}
\input{5_experiments}
\input{6_conclusions}


\section*{Acknowledgments}
This work has been partly supported by the Horizon Europe Programme under GA.101135513 (CyclOps) and the Spanish Ministerio de Ciencia e Innovación under project PID2023-152841OA-I00 / AEI/10.13039/501100011033 (TALC). Anna Queralt is a Serra Húnter Fellow.

\bibliographystyle{IEEEtran}
\bibliography{references}  

\begin{IEEEbiography}[{\includegraphics[width=1in,height=1.25in,clip,keepaspectratio]{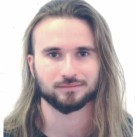}}]{Marc Maynou}
received his MSc degree in Data Science from Universitat Politècnica de Catalunya, Barcelona, Spain in 2023. He is currently working toward his PhD degree with the Database Technologies and Information Management group, Barcelona, Spain, supervised by Prof. Sergi Nadal. His research interests include data discovery, big data management and table representation learning.
\end{IEEEbiography}

\begin{IEEEbiography}[{\includegraphics[width=1in,height=1.25in,clip,keepaspectratio]{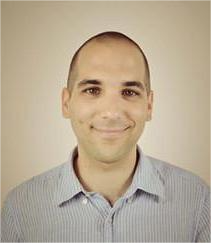}}]{Sergi Nadal}
received the joint Ph.D. degree in computer science from the Universitat Politècnica de Catalunya (UPC) and the Université Libre de Bruxelles (ULB) in 2019. He is an Associate Professor with the Department of Service and Information System Engineering at UPC, and a member of the Database Technologies and Information Management (DTIM) research group. His research interests lie in the area of data and information management, with a focus on Data Integration, Data Discovery, and the automation of the end-to-end data lifecycle.
\end{IEEEbiography}

\begin{IEEEbiography}[{\includegraphics[width=1in,height=1.25in,clip,keepaspectratio]{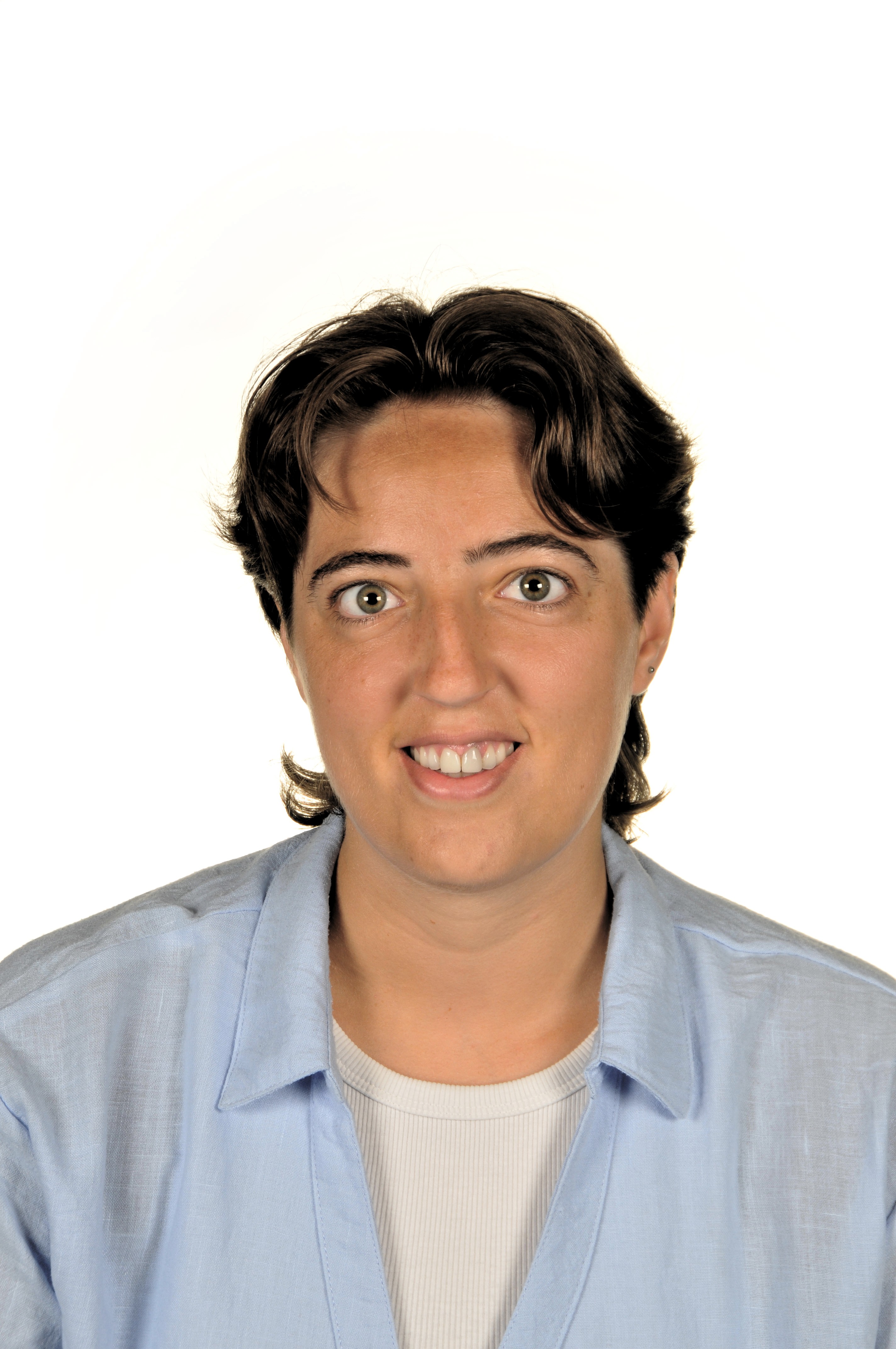}}]{Raquel Panadero}
holds a degree in Telecommunications Engineering from the Universitat Politècnica de Catalunya and is currently completing her Master’s in Data Science at TU Wien. Alongside her studies, she worked as a Researcher at the Computer Vision Lab (TU Wien), focusing on video, emotion, and action recognition models. Currently, Raquel is an AI Engineer at Big Blue Marble, where she develops innovative AI-driven solutions for the audiovisual and media industry.
\end{IEEEbiography}

\begin{IEEEbiography}[{\includegraphics[width=1in,height=1.25in,clip,keepaspectratio]{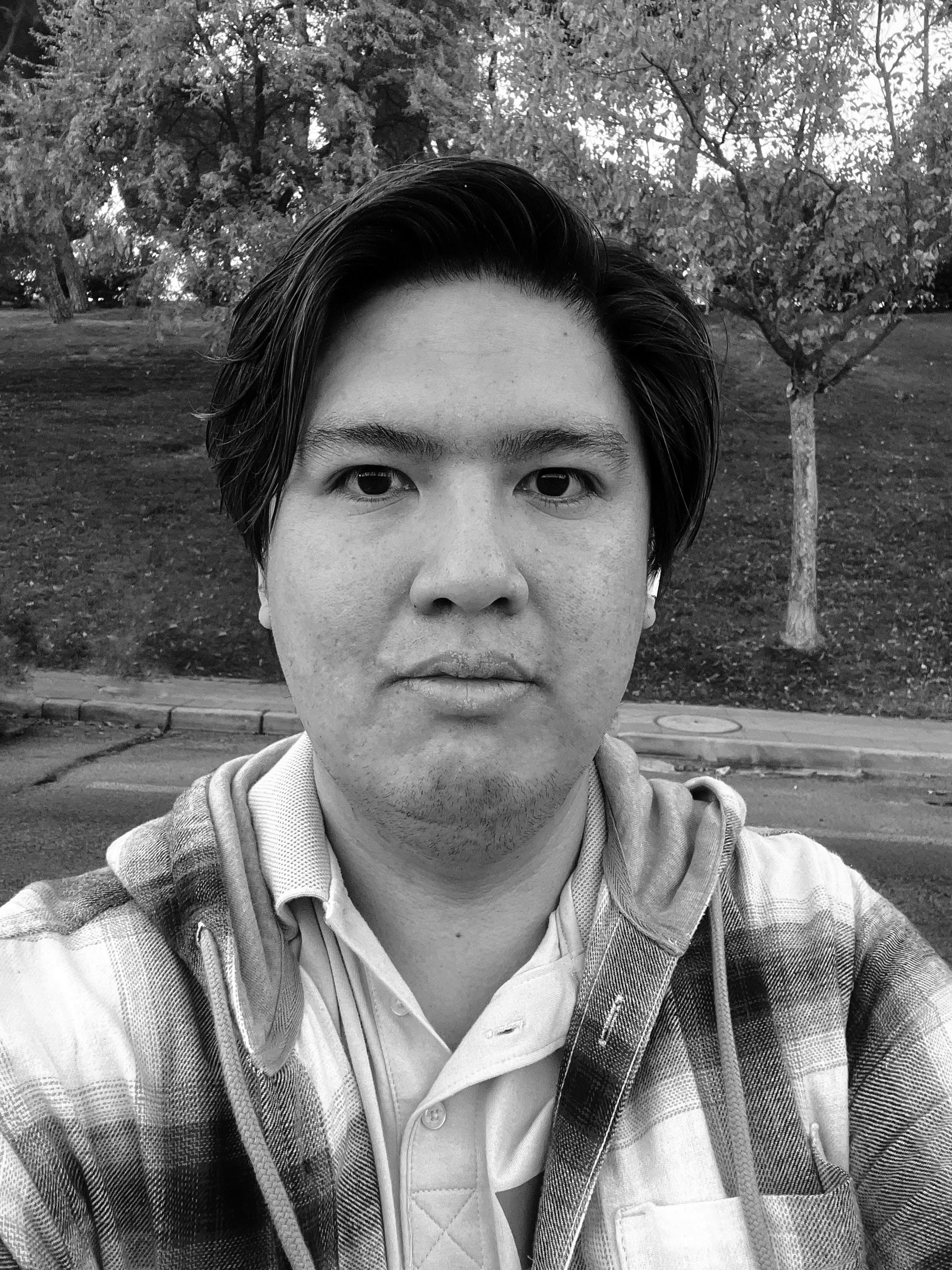}}]{Javier Flores}
received the PhD degree in computer science from Universitat Politècnica de Catalunya (UPC) in 2025. His research focuses on scalable and automated data integration, with an emphasis on graph-based schemas and data discovery techniques for heterogeneous data sources.
\end{IEEEbiography}

\begin{IEEEbiography}[{\includegraphics[width=1in,height=1.25in,clip,keepaspectratio]{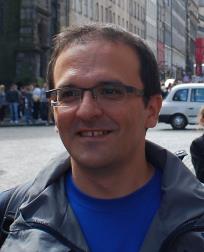}}]{Oscar Romero}
is a full professor at Universitat Politècnica de Catalunya, BarcelonaTech (UPC) and a member of the DTIM research group. His research focuses on data-intensive systems with the ultimate objective to operationalize and automate different stages of the complete data lifecycle in fields such as Big Data, Data Science and data-driven Artificial Intelligence. He has published 100+ papers on top venues and served in the Programme Committee of top-tier conferences such as VLDB, ICDE, EDBT, ISWC, WWW, IEEE Big Data, as well as successfully supervised 9 PhD theses. 
\end{IEEEbiography}

\begin{IEEEbiography}[{\includegraphics[width=1in,height=1.25in,clip,keepaspectratio]{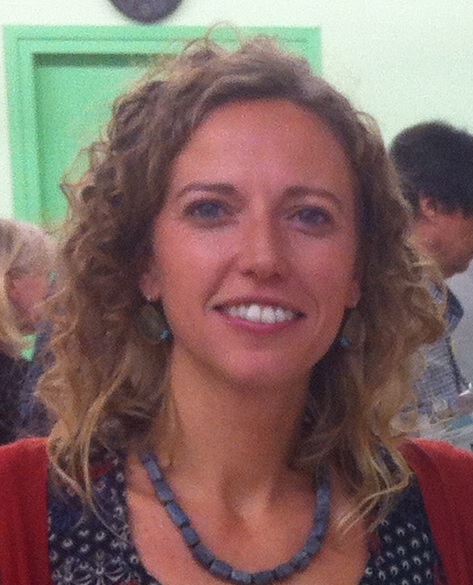}}]{Anna Queralt}
is an Associate Professor and Serra Húnter Fellow at Universitat Politècnica de Catalunya (UPC), which she joined in 2021. She leads the Database Technologies and Information Management (DTIM) research group at UPC, and coordinates the Master in Data Science at Facultat d'Informàtica de Barcelona. She was a Senior Researcher at the Barcelona Supercomputing Center from 2012 to 2024, leading the Distributed Object Management research line. She has led and participated in numerous European and national projects, as well as industrial contracts in the topics of her research. Her research interests are related to data management and processing, aiming at the automation of the different steps of the data lifecycle, from data modeling to processing and sharing.
\end{IEEEbiography}

\end{document}

%% file: 1_introduction.tex
\section{Introduction} \label{sec:intro}


Data discovery involves navigating vast data sources to find relevant datasets and their relationships \cite{DBLP:conf/icde/FernandezMQEIMO18, BogatuFP020}, making it a vital step in the Data Science pipeline for applications like search and data augmentation \cite{DBLP:conf/sigmod/Fan00M23}. This is particularly important for data-driven organizations, which rely on \textit{data lakes} \cite{DBLP:journals/tkde/HaiKQJ23,DBLP:journals/pvldb/NargesianZMPA19}, where they centralize the storage of their data assets (i.e., any dataset potentially relevant to the organization's data analysts) to promote accessibility and reusability. 


Data discovery on data lakes presents several challenges. First, data lakes are characterized by \textbf{data heterogeneity}, including semantic diversity (disparate topics) and syntactic variability (differences in attributes, values or cardinalities) \cite{DBLP:journals/tkde/HaiKQJ23}. Second, data lakes contain a \textbf{massive number of datasets}, typically not large in cardinality but with a substantial number of columns. These two characteristics arise from the fact that data lakes store data coming from a myriad of independent stakeholders, whose generated datasets present properties unique to the domain they belong to. These distinctive features render classical data discovery solutions that rely on integrity constraints (e.g. PK-FKs) infeasible and ineffective. Thus, in sight of the above-mentioned challenges, we study the problem of data discovery tasks in data lakes. More precisely, we focus on computing rankings of joinable attributes. 
The problem is defined as follows: given a \textit{query column}, our aim is to discover\textit{ high-quality joins} with columns in the data lake, where a high-quality join produces a semantically rich dataset. Materializing these joins provides downstream applications with an extended set of features from the data lake.


State-of-the-art join discovery methods can be classified as either \textit{syntactic} or \textit{semantic}. 
Syntactic approaches define joinability via set-overlap metrics (e.g. Jaccard or containment), following the traditional inner-join paradigm. Joins are thus ranked higher in proportion to the number of shared values between two columns
Since performing pairwise column computations is unfeasible on a data lake, these systems propose mechanisms to approximate their calculation (e.g. locality-sensitive hashing schemes \cite{DBLP:journals/pvldb/ZhuNPM16,DBLP:conf/icde/FernandezAKYMS18}, or compact inverted indexes \cite{DBLP:journals/pvldb/DengKMS17,DBLP:conf/sigmod/ZhuDNM19}). 
While these methods are scalable, the quality of the generated rankings is negatively affected by data heterogeneity \cite{DBLP:conf/edbt/0002N021}. 
As illustrated in Example 1, when the same values are used to represent semantically disparate concepts, value-overlap methods are prone to false positives.

\begin{examplebox}{Example 1}
Consider the toy datasets of Figure \ref{fig:toyDatasets}. Using $D_1.Country$ as query column, a purely syntactic approach based on value overlap would identify $D_2.Country$, $D_3.Language$ and $D_4.X$ as good candidates.
The first observation is that schema-based methods, such as LogMap \cite{DBLP:conf/semweb/Jimenez-RuizG11}, would fail to propose the combination $D_1.Country = D_4.X$ due to the lack of embedded semantics in the schema of $D_4$. 
More importantly, the pair $D_1.Country = D_3.Language$ constitutes a false positive; despite value overlap, they represent different semantic concepts, and their join would be meaningless.
\end{examplebox}

\begin{figure*}[t!]
    \centering
    \begin{minipage}[t]{0.2\textwidth}
        \centering
        \small
        \caption*{$D_1$ -- Happiness scores}
        \begin{tabular}{|c|c|c|}
            \hline
            \textbf{Country} & \textbf{Happiness}\\ \hline
            POL & 6.595 \\ \hline
            DEU  & 6.354 \\ \hline
            ITA & 6.892 \\ \hline
            JPN & 6.592 \\ \hline
        \end{tabular}
    \end{minipage}%
    \hspace{0.02\textwidth} 
    \begin{minipage}[t]{0.3\textwidth}
        \centering
        \small
        \caption*{$D_2$ -- Satisfaction index}
        \begin{tabular}{|c|c|c|c|c|}
            \hline
            \textbf{Country} & \textbf{Location} & \textbf{Satisfaction} \\ \hline
            ITA & Bologna & 1484,7 \\ \hline
            ITA & Milan & 1526,7 \\ \hline
            DEU & Munich & 1678,6  \\ \hline
            DEU & Berlin & 1720,6 \\ \hline
        \end{tabular}
    \end{minipage}%
    \hspace{0.02\textwidth} 
    \begin{minipage}[t]{0.2\textwidth}
        \centering
        \small
        \caption*{$D_3$ -- Language speakers}
        \begin{tabular}{|c|c|c|}
            \hline
            \textbf{Language} & \textbf{Speakers}\\ \hline
            DEU & 134M \\ \hline
            POL & 50M \\ \hline
            ITA & 67M \\ \hline
            JPN & 123M \\ \hline
        \end{tabular}
    \end{minipage}
    \hspace{0.02\textwidth}
    \begin{minipage}[t]{0.2\textwidth}
        \centering
        \small
        \caption*{$D_4$ -- Population data}
        \begin{tabular}{|c|c|c|}
            \hline
            \textbf{X} & \textbf{Y} & \textbf{Z} \\ \hline
            DEU & 84.7M & 2024 \\ \hline
            ITA & 58.9M & 2024 \\ \hline
            POL & 37.5M & 2024 \\ \hline
            JPN & 123.7M & 2024 \\ \hline
        \end{tabular}
    \end{minipage}
    \caption{\label{fig:toyDatasets} Toy datasets}
\end{figure*}

\IEEEpubidadjcol


In contrast to syntactic approaches, semantic methodologies enhance the accuracy of join discovery by assessing relatedness in a higher dimensional space. To that end, Table Representation Learning (TRL) techniques are used to generate embeddings that capture the semantic properties of columns. 
These systems leverage deep learning algorithms that fine-tune pre-trained language models \cite{BogatuFP020, fan2022semantics} or general-purpose knowledge bases \cite{khatiwada2023santos} to semantically categorize columns. These approaches tend to produce highly accurate rankings of candidate joins, yet they are resource-intensive processes that require high-end GPUs and large memories to compute and store the necessary artifacts. 

Overall, syntactic approaches can operate in large-scale contexts since their methods are highly scalable, but they generate a considerable amount of false positives due to not properly addressing data heterogeneity. Conversely, TRL-based methods improve joinability assessment by capturing semantic relationships, but they are constrained by high execution costs from the underlying deep learning methods.

\subsection{\sys:~effective and efficient joinability discovery}

In order to overcome the limitations present in the state-of-the-art, we present \sys, a novel approach to accurately and efficiently compute rankings of joinable columns on data lakes. Our approach aims to retain the best of both syntactic and semantic families. \sys~enjoys the simplicity and efficiency of approximating the computation of a syntactic metric while including mechanisms to capture the semantic relationships of the sets of values to improve joinability assessment. \sys~introduces the following two novelties.

\medskip

\subsubsection{A novel, semantic-aware data lake joinability metric} 
Traditional set-overlap metrics are prone to false positives as they lack a mechanism to detect semantic relatedness. Hence, we propose an alternative to classical approaches that avoids costly embedding representations by introducing semantic-awareness into a scalable syntactic metric.
We first employ a \textbf{multiset variant of the Jaccard} index for set-overlap measurement, addressing the overrepresentation of multisets in data lakes \cite{DBLP:journals/corr/abs-2110-09619}. Second, building on research showing that set-overlap metrics are inaccurate when column cardinalities differ greatly \cite{DBLP:journals/pvldb/NaziDNC18}, which is common in real-world data lakes due to data variability and diversity, we use the \textbf{cardinality proportion} between sets as a lightweight proxy for semantic comparison. 
Our final metric is a linear combination of these two measurements, and, as we demonstrate experimentally in Section \ref{sec:exp_effectiveness}, the quality of the generated rankings is comparable to that obtained by TRL-based approaches.

\medskip

\subsubsection{A lightweight structure to approximate the joinability metric} Computing our novel metric is still expensive in large-scale data lakes due to the required set-overlap evaluation. We, therefore, propose to predict it based on \textbf{data profiles}, compact descriptive representations of a column. Underpinning our method is the hypothesis that profile similarity is related to the semantic relatedness of columns, assuming that a generalized closeness in the properties included in the profiles identifies meaningfully related columns. For that, we rely on state-of-the-art data profiling techniques \cite{DBLP:journals/vldb/AbedjanGN15} to capture profiles related to value distributions, syntactic properties and cardinalities. Profile computation is highly scalable, reducing preparation time by orders of magnitude and minimizing hardware demands. The model we train is general, lightweight, and requires no fine-tuning for specific environments. We showcase the scalability improvements in Section \ref{sec:exp_efficiency}.

\stitle{Contributions} Our contributions are as follows:

\begin{itemize}[leftmargin=*]
    \item We introduce a novel, semantic-aware metric to determine the join quality of two columns in data lakes, considering their multiset Jaccard index and cardinality proportion.
    \item We learn a highly accurate, general, and scalable model to predict the abovementioned metric, trained via profiles.
    \item We develop a system that combines both of these notions into a cohesive tool to perform data discovery tasks at scale.
\end{itemize}

\noindent\textbf{Outline.}
The remainder of this paper is organized as follows. Section \ref{sec:relatedwork} reviews related work. Sections \ref{sec:preliminaries}-\ref{sec:predictingQ} define our join discovery metric and the model to predict it. Section \ref{sec:experiments} evaluates our approach's effectiveness and scalability, and Section \ref{sec:conclusions} concludes the paper.

%% file: 2_relatedwork.tex
\section{Related Work}\label{sec:relatedwork}

\renewcommand{\arraystretch}{1.2}

In this section, we review related work on join discovery. Classical solutions relying on integrity constraints (e.g., PK-FKs~\cite{jiang2020holistic}) are ineffective in data lakes due to their heterogeneous and large-scale data. Hence, we focus on modern solutions, distinguishing between \textbf{syntactic} methods that approximate set-overlap metrics and \textbf{semantic} ones based on TRL. We overview these systems in Table \ref{tab:sota_class}.

\begin{table}[H]
\centering
\caption{Join discovery methods classification\label{tab:sota_class}}
\vspace{3mm}
\resizebox{1\columnwidth}{!} {
    \begin{minipage}{\columnwidth}
        \centering  
        \small
        \begin{tabular}{|c|c|}
            \hline
            \multicolumn{2}{|c|}{\textbf{Similarity assessment}} \\ 
            \multicolumn{2}{|c|}{Semantic \dotfill Syntactic} \\
            \hline
            Table Representation Learning & \phantom{........} Column-overlap \phantom{........} \\ 
            \cite{khatiwada2023santos, fan2022semantics, dong2022deepjoin, DBLP:journals/pvldb/BharadwajGBG21, DBLP:journals/corr/abs-2212-14155, DBLP:conf/icde/DongT0O21, BogatuFP020} & \cite{DBLP:conf/icde/FernandezAKYMS18, DBLP:journals/pvldb/ZhuNPM16, DBLP:conf/sequences/Broder97, DBLP:journals/tods/XiaoWLYW11, DBLP:conf/sigmod/ZhuDNM19, DBLP:journals/pvldb/DengKMS17} \\ 
            \hline
            \multicolumn{2}{|c|}{Expensive \dotfill Efficient} \\
            \multicolumn{2}{|c|}{\textbf{Algorithmic complexity}} \\ 
            \hline
        \end{tabular}
    \end{minipage}
}
\end{table}

\stitle{Syntactic approaches} 
\textbf{SilkMoth}~\cite{DBLP:journals/pvldb/DengKMS17} generates token-based signatures to prune the search space with an inverted index. \textbf{JOSIE} \cite{DBLP:conf/sigmod/ZhuDNM19} and \textbf{PPJoin} \cite{DBLP:journals/tods/XiaoWLYW11} optimize the number of comparisons when computing the similarity metric. JOSIE builds a ranked list from the $k$ candidate tables with highest containment. PPJoin employs prefix filtering to avoid computing similarity values for all possible values. 
\textbf{MinHash} \cite{DBLP:conf/sequences/Broder97} uses the minwise hash function, with a collision probability equal to the Jaccard similarity, requiring, for every value, to compute the MinHash signature hundreds of times. \textbf{LSH Ensemble} \cite{DBLP:journals/pvldb/ZhuNPM16} focuses on finding attributes with a high containment similarity. 
\textbf{Aurum} \cite{DBLP:conf/icde/FernandezAKYMS18} represents relations between datasets and their attributes in a graph data structure. In this graph, attribute nodes are related if their hashing signatures, generated from their instances, are similar in an LSH index.

\stitle{Semantic approaches}
\textbf{D3L} \cite{BogatuFP020} employs LSH indexes generated from four features plus word embeddings, defining joinability on their composition. \textbf{PEXESO} \cite{DBLP:conf/icde/DongT0O21} proposes an approach based entirely on embeddings with partitioning techniques to alleviate indexing costs. \textbf{WarpGate} \cite{DBLP:journals/corr/abs-2212-14155} also applies embeddings and leverages LSH indexes for fast similarity searches. Some systems train specialized models: \textbf{DLN}~\cite{DBLP:journals/pvldb/BharadwajGBG21} uses metadata features for Microsoft's Cosmos data lake, and \textbf{DeepJoin}~\cite{dong2022deepjoin} fine-tunes a BERT-based model for join detection. 
\textbf{Starmie} \cite{fan2022semantics} advances this with contrastive learning to train column encoders, enriching semantics with table context. Others leverage knowledge bases (KBs): \textbf{SANTOS}~\cite{khatiwada2023santos} maps semantics to columns from a KB, \revision{and \textbf{KGLiDS}~\cite{helali2024kglids} advances the paradigm by providing a unified semantic layer that integrates knowledge graphs, data representations (column embeddings and profiles) and automation pipelines}.

\stitle{Research gap} 
Semantic approaches are the state-of-the-art in accuracy, yet TRL-based methods demand extensive preparation time and computational resources. \revision{This gap motivates \sys, a new paradigm for join discovery achieving high effectiveness through an efficiency-first design. We introduce a scalable metric that combines multiset Jaccard and cardinality proportion to deliberately avoid costly embeddings. We then infer this metric using a generic, lightweight model trained on data profiles. This design ensures our method is scalable and handles new data without requiring fine-tuning.}


%% file: 3_scalable_join_discovery.tex
\section{A semantic-aware joinability metric}\label{sec:preliminaries}

\noindent This section provides background for the proposed method. We discuss limitations of column-overlap metrics for ranking joinable pairs, and introduce the cardinality proportion's role.

\subsection{Join quality: a metric of semantic relatedness}\label{sec:problem_definition}

\noindent Let $\mathcal{C}$ be the set of all columns in a data lake, a \textbf{join quality metric} is a function $\mathcal{Q}: (A, B) \rightarrow \mathbb{R}_{[0,1]}$, such that $A,B \in \mathcal{C}$. Furthermore, for any $A,B,C \in \mathcal{C}$, it holds that $\mathcal{Q}(A,B) > \mathcal{Q}(A,C)$ if $\langle A,B \rangle$ yields a higher quality join than $\langle A,C \rangle$. Intuitively, a join quality metric ranks potential joins between attributes based on their presumed quality, employing some pre-determined criteria to do so. 

Measuring semantic closeness is complex, often solved by computing embedding similarity. In contrast to TRL methods, we aim to reduce complexity by using a qualitative assessment encompassing two join typologies. 
A pair $\langle A,B \rangle$ is a \textbf{syntactic join} if $A \cap B \neq \emptyset$. Following \cite{DBLP:conf/icde/KoutrasSIPBFLBK21}, $\langle A,B \rangle$ is a \textbf{semantic join} if it is syntactic and a function $h: A \leftrightarrow  B$ exists denoting semantic equivalence (i.e. both refer to the same concept). Thus, for a semantic-aware metric $\mathcal{Q}$, it holds that $\mathcal{Q}(A,B) > \mathcal{Q}(A,C)$ if $\langle A,B \rangle$ is a semantic join and $\langle A,C \rangle$ is not.

\begin{examplebox}{Example 2}
In Figure \ref{fig:toyDatasets}, both $\langle D_1.Country, $ $D_2.Country \rangle$ and $\langle D_1.Country, $ $D_3.Language \rangle$ are syntactic joins, as their intersection is non-null. Yet, only the former is a semantic join, as both columns refer to countries. A join between $D_1.Country$ and $D_3.Language$ would be semantically meaningless and lead to misleading interpretations.
\end{examplebox}

Prioritizing semantic joins over syntactic ones favors meaningful relationships over a high number of matches. Thus, a low-overlapping semantic join is preferable to a high-overlapping syntactic join, as the combined information will be significant even with fewer matches.


\subsection{Why column-overlap metrics are not sufficient}\label{sec:multiset_jaccard}

Data lakes often contain fully denormalized datasets where PK-FK constraints cannot be assumed. Thus, repeated values (i.e. multisets) pose a problem, as they can lead to inflated overlap scores that misrepresent the true relationship between datasets. Neither containment ($C(A,B) = \frac{ |A \cap B|}{|A|}$) nor Jaccard ($J(A,B) = \frac{ |A \cap B|}{|A \cup B|}$), the two most popular syntactic metrics, address this issue. In both cases, repeated values can bloat the score and mislead the perceived similarity. 
In contrast, the community has proposed overlap metrics that account for repeated elements within a set. Precisely, variations of the Jaccard index have been defined \cite{DBLP:journals/corr/abs-2110-09619}, such as the multiset Jaccard index $\mathcal{J}(A,B) = \frac{ |A \cap B|}{|A| + |B|}$ (note that its scores are bounded in the range [0, 0.5]).




To determine if column-overlap metrics can compute accurate rankings of semantically-joinable pairs, we built the \sys$_{\text{BM}}$ data lake\footnote{Such ground truth is publicly available in the paper's companion \href{https://freyja-data-discovery.github.io/}{website}} from 160 datasets from open repositories, whose domains vary widely. This yielded 4,318 candidate joins (with $\geq 10\%$ overlap), which we manually labeled as semantic (1,696) or syntactic (2,622). Next, following \cite{DBLP:journals/pvldb/CongHSGJ23}, we computed baseline quality rankings for a sample of query columns employing cosine similarity over their embeddings. We consider this TRL-based ranking as the ground truth, given the superior performance of such approaches.

\begin{figure}[h]
	\begin{center}
		\includegraphics[width=0.9\linewidth]{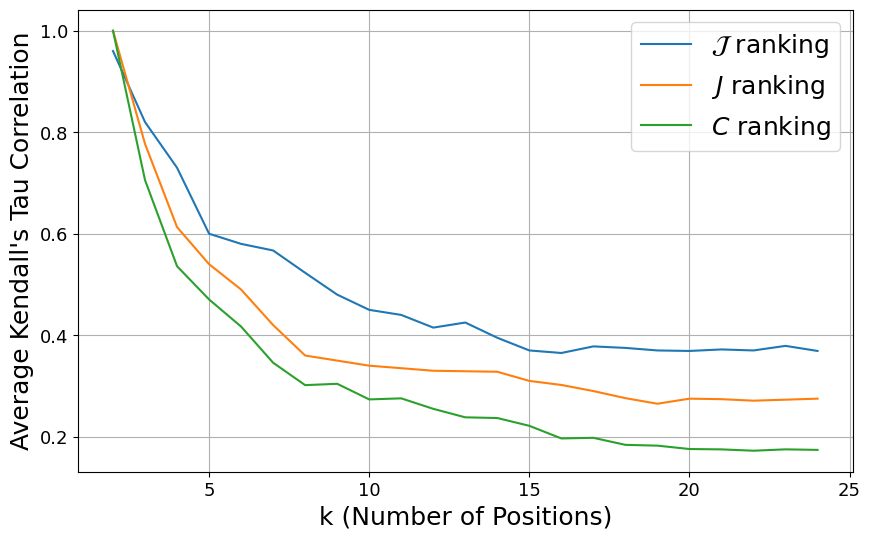}
		\caption{Average Kendall's $\tau$ coefficient between embedding-based quality rankings and set-overlap metrics}
		\label{fig:ranking_quality}
	\end{center}
\end{figure}

To evaluate column-overlap metrics, we then generated quality rankings using $C$, $J$, and $\mathcal{J}$. Figure \ref{fig:ranking_quality} depicts the average Kendall's rank correlation coefficient between each of these rankings and the embedding-based baseline (BERT) for an increasing ranking size ($k$). We observe that the ability of these metrics to produce high-quality rankings rapidly decreases. This trend held when using other embedding models (RoBERTa, TaBERT) and other ranking correlation statistics (Spearman's, Normalized Discounted Cumulative Gain).



Figure \ref{fig:ranking_quality} shows that column-overlap metrics are ill-suited for ranking semantically joinable pairs, as they assess relatedness purely on shared values. Since data lakes exhibit high semantic and syntactic heterogeneity, join detection requires a mechanism to capture relatedness beyond simple set intersection.

\subsection{On the role of the cardinality proportion}

Previous research shows set-overlap metrics are inaccurate when the cardinality of a column is comparatively larger than the other's \cite{DBLP:journals/pvldb/NaziDNC18}. Figure \ref{fig:ranking_quality} confirms that multiset Jaccard is slightly superior as it addresses multisets by considering cardinalities in isolation. We thus propose that in data lakes, large cardinality differences can serve as a lightweight proxy for semantic dissimilarity. We define our experimental hypothesis as: \textit{\textbf{two sets of values with widely different cardinalities represent different semantic concepts}}. \revision{Our hypothesis is grounded in the concept of \textit{Natural Kinds} \cite{quine1969natural}, where distinct semantic concepts (e.g., countries) have a characteristic domain size. In a data lake, a column's cardinality acts as a lightweight semantic signature; therefore, a large difference in cardinality is a strong indicator that two columns represent different concepts.} 
Below, we exemplify this intuition.

\begin{examplebox}{Example 3}
We previously stated the join $\langle D_1.Country,$ $D_3.Language \rangle$ was syntactic. The underlying datasets represent domains with vastly different sizes: there are ~200 countries but over 7,100 languages. Following our hypothesis, this large difference in the number of values is itself indicative that the two sets represent different concepts.
\end{examplebox}


\noindent Comparing the cardinalities of two sets of data can be summarized by the cardinality proportion, $K(A,B) = \frac{ min(|A|,|B|)}{max(|A|,|B|)}$. \revision{We motivate our intuition from an information theory perspective. A data column's maximum possible entropy is monotonically related to its cardinality: $H_{\text{max}}(A) = \log_{2}(|A|)$. If two columns are semantically equivalent, they are drawn from the same information source and should have a nearly identical capacity to encode information. This implies their maximum entropies, $H_{max}(A)$ and $H_{max}(B)$, are nearly identical which in turn means their cardinalities must also be very close. Conversely, a large discrepancy in cardinality implies a large discrepancy in information content. If $|A|\gg|B|$, it follows that $H_{max} (A)\gg H_{max} (B)$. It is, hence, information-theoretically improbable that two concepts with vastly different levels of complexity, granularity or scope could be semantically equivalent.}

\begin{figure}
	\begin{center}
		\includegraphics[width=0.97\linewidth]{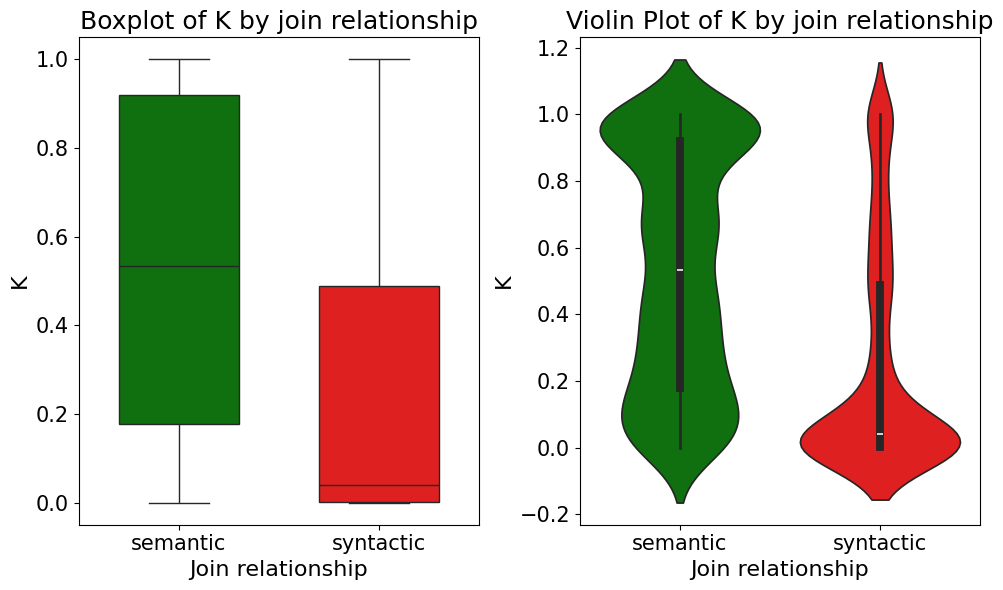}
		\caption{\revision{Distribution of the cardinality proportion ($K$) for semantic vs. syntactic joins on the \sys$_{\text{BM}}$ benchmark}}
		\label{fig:violin_plot}
	\end{center}
\end{figure}

\stitle{Empirical validation} \revision{To validate our hypothesis, we analyzed the distribution of $K$ across join types for the \sys$_{\text{BM}}$ benchmark (Figure \ref{fig:violin_plot}) as well as those in Section \ref{sec:experiments} (to be found in the companion website). The results confirm our hypothesis: semantic joins consistently exhibit significantly higher $K$ values, with their distributions concentrated at the upper end of the scale. However, the analysis also reveals some overlap in the distributions, indicating that $K$ alone is insufficient for a perfect assessment of join quality.} 

The use of $K$ to detect PK-FK constraints is common in relational databases \cite{rostin2009machine, zhang2010multi}, but it has not been studied for join discovery. From our findings, we define a join quality metric that combines a column-overlap metric with the cardinality proportion. We choose multiset Jaccard as it empirically outperforms others. Combining $\mathcal{J}$ and $K$ yields a metric that merges a syntactic (overlap) assessment with a lightweight semantic characterization. In Section \ref{sec:join_quality_metric} we explore how to combine the two measurements into a cohesive metric. 

%% file: 4_join_quality_metric.tex
\section{Join Quality Metric}\label{sec:join_quality_metric}

To combine multiset Jaccard ($\mathcal{J}$) and cardinality proportion ($K$) into a single metric, we first define a discrete, multi-class metric that assigns joins to one of $L$ quality buckets, and then generalize it into a continuous quality assessment.

\subsection{Discrete quality metric}\label{sec:discrete_metric}

\begin{figure}
    \begin{center}
        \includegraphics[width=.85\linewidth]{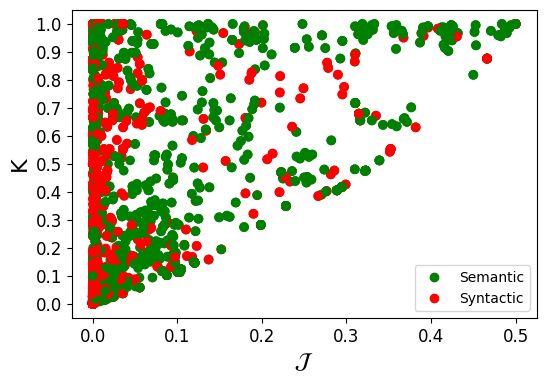}
        \caption{Distribution of ground truth labels over $\mathcal{J}$ and $K$}
        \label{fig:gt_scatter}
    \end{center}
\end{figure}

Many possible combinations of $\mathcal{J}$ and $K$ exist. A simple empirical approach, such as defining a function to separate semantic from syntactic joins in our ground truth, is unfeasible. As shown in Figure \ref{fig:gt_scatter}, the distributions for both join types overlap significantly, meaning any single separation function would overfit to our data and fail to generalize. Therefore, a more general approach adaptable to unseen data is required.

We first generate a discrete metric that divides the join space into $L$ quality buckets, each representing a different degree of joinability. Each bucket corresponds to a pre-defined range of $\mathcal{J}$ and $K$, where higher values of both indicate a higher quality join. As evidenced in Figure \ref{fig:gt_scatter}, $\mathcal{J}$ provides a stronger initial quality assessment (most syntactic joins have $\mathcal{J} < 0.05$), while $K$ acts as a complementary semantic-aware constraint. Thus, we define a function $Q(A,B,L)$ assigning joins to $L$ buckets based on $\mathcal{J}(A,B)$ and constrained by $K(A,B)$.


\begin{equation*}
    \resizebox{\hsize}{!}{$Q(A,B,L) = max(i) \in [0, ..., L] | \mathcal{J}(A,B) \geq \dfrac{1}{2^{i+1}} \wedge K(A,B) \geq (1-\dfrac{i}{L})$}
\end{equation*}

\begin{figure}
	\begin{center}
		\includegraphics[width=.85\linewidth]{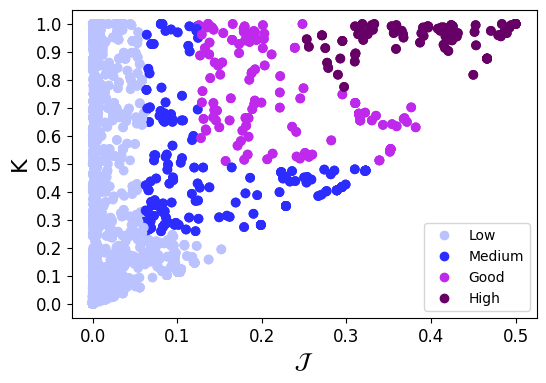}
		\caption{Distribution of $Q(A,B,L)$ for the case of $L=4$}
		\label{fig:gt_quality_levels}
	\end{center}
\end{figure}

For illustration, the results of applying $Q(A,B,4)$ to \sysbm~are shown in Figure \ref{fig:gt_quality_levels}. \revision{The formulation of this function is based on the principle of prioritizing  $\mathcal{J}$ as the primary indicator of join quality, while leveraging $K$ as a semantic relatedness constraint. This design imposes exponentially decreasing thresholds for J and linearly decreasing thresholds for K across the quality buckets.} Intuitively, higher quality buckets will contain a larger proportion of semantic joins. This function is agnostic of the underlying data and can be applied to any data lake. This multi-class separation provides a generalized join assessment based on our assumption that higher $\mathcal{J}$ and $K$ values correlate with semantic joins, without overfitting to any specific dataset. We empirically demonstrate its generalizability in Section \ref{sec:experiments}.

\subsection{Continuous quality metric}\label{sec:continuous_metric}

While $Q(A,B,L)$ provides quality levels beyond binary labels, its discrete output limits the granularity of rankings as they are bounded by $L$. To overcome this, we generalize the discrete metric into a continuous score in the range $[0,1]$. 
Our approach is to fit a continuous probability distribution to the cumulative distribution function (CDF) of $Q(A,B,L)$ for a given $L$. We use the Gaussian normal distribution for this task due to its flexibility in modeling natural phenomena. Specifically, we use a multivariate normal distribution for $\mathcal{J}$ and $K$. Assuming variable independence, the truncated two-dimensional CDF, $cdf_{\mathcal{J},K} \left(j,k\right)$, is the product of the individual CDFs:

\begin{equation*}
    \resizebox{\linewidth}{!}{$ 
        cdf_{\mathrm{\mathcal{J},K}} \left(j,k\right)=\frac{\Phi \left(\frac{j-\mu_{j} }{\sigma_{j} }\right)-\Phi \left(\frac{a-\mu_{j} }{\sigma_{j} }\right)}{\Phi \left(\frac{b-\mu_{j} }{\sigma_{j} }\right)-\Phi \left(\frac{a-\mu_{j} }{\sigma_{j} }\right)}\frac{\Phi \left(\frac{k-\mu_k }{\sigma_k }\right)-\Phi \left(\frac{a-\mu_k }{\sigma_k }\right)}{\Phi \left(\frac{b-\mu_k }{\sigma_k }\right)-\Phi \left(\frac{a-\mu_k }{\sigma_k }\right)}
    $}
\end{equation*}

\noindent where $\Phi(x)$ is the univariate CDF of the normal distribution defined as:

\begin{equation*}
\Phi \left(x\right)=\frac{1}{2}\left(1+\mathrm{erf}\left(\frac{x}{2}\right)\right)=\frac{1}{2}\left(1+\frac{2}{\sqrt{\pi }}\int_0^{\frac{x}{2}} e^{-t^2 } \mathrm{dt}\right)
\end{equation*}

\medskip

The challenge is to find the mean values ($\mu_{j}$, $\mu_k$) and the covariance matrix $\Sigma$ that best fit the discrete function $Q(A,B,L)$. To do this, we exhaustively search for the parameters that minimize the Wasserstein distance between the discrete and continuous distributions. Applied to \sysbm, this process yielded the following results: 
\begin{equation*}
\mu_{mj} = 0.44, \mu_k = 0, \Sigma = \begin{pmatrix}
0.19 & 0\\ 
0 & 0.28
\end{pmatrix}
\end{equation*}

\noindent Figure \ref{fig:cdfs} shows the fitted normal CDF against the empirical distribution function (EDF) of $Q(A,B,4)$. \revision{We chose $L=4$ after testing for $2 \leq L \leq 16$, as it minimized the distance for both $\mathcal{J}$ and $K$ while balancing bucket quantity and interpretability.} The resulting metric is defined as:

\begin{figure}
	\centering
	\begin{minipage}{0.5\linewidth}
		\centering
		\includegraphics[width=1\linewidth]{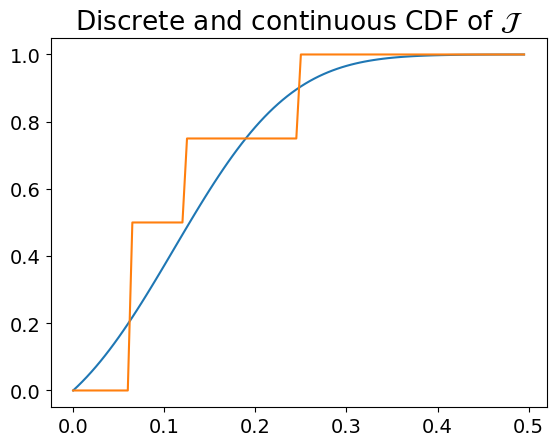} 
	\end{minipage}\hfill
	\begin{minipage}{0.5\linewidth}
		\centering
		\includegraphics[width=1\linewidth]{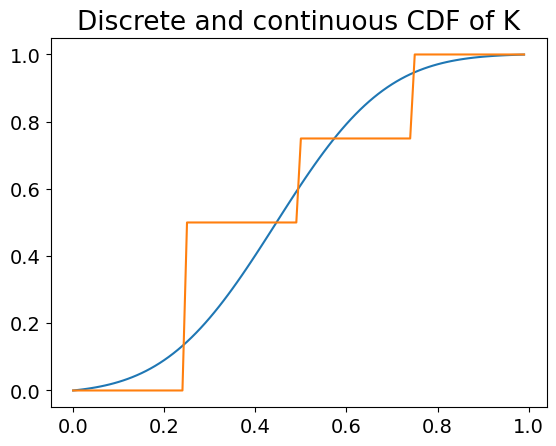} 
	\end{minipage}
	\caption{Resulting \textit{cdf}s that minimize the Wasserstein distance over the \textit{edf} of $\mathcal{J}$ (left) and $K$ (right) for $Q(A,B,4)$}\label{fig:cdfs}
\end{figure}

\begin{equation*}
    \resizebox{\linewidth}{!}{$ 
        cdf \Big( \mu_{\mathcal{J}},\Sigma[0][0],0,0.5, \mathcal{J}(A,B) \Big) cdf \Big( \mu_K,\Sigma[1][1],0,1,K(A,B) \Big)
    $}
\end{equation*}




\section{Predicting the join quality} \label{sec:predictingQ}

The metric from Section \ref{sec:join_quality_metric}, while effective, is costly to compute due to its reliance on set intersection. To ensure scalability while maintaining the effectiveness of semantic approaches, we employ a \textbf{learning approach} to predict join quality using \textbf{data profiles}.

\subsection{Profile-based join quality prediction}

Predicting join quality via a learning model is more efficient than direct computation, as inference is faster than calculating intersections. Since raw column values are too large and heterogeneous for a model, we require a summarized and consistent data representation as input. We use data profiles for this task. A profile $P(A)$ for a column $A$ is a set of meta-features $\{ m_1, \ldots, m_n \}$ summarizing its content and structure (e.g. cardinality, entropy, average length). Profiles provide a compact representation of the original column's characteristics, and offer a high-level understanding of a column's properties without manual data inspection. Their uniform structure makes them ideal for training and inference.

\begin{figure}
	\begin{center}
		\includegraphics[width=\linewidth]{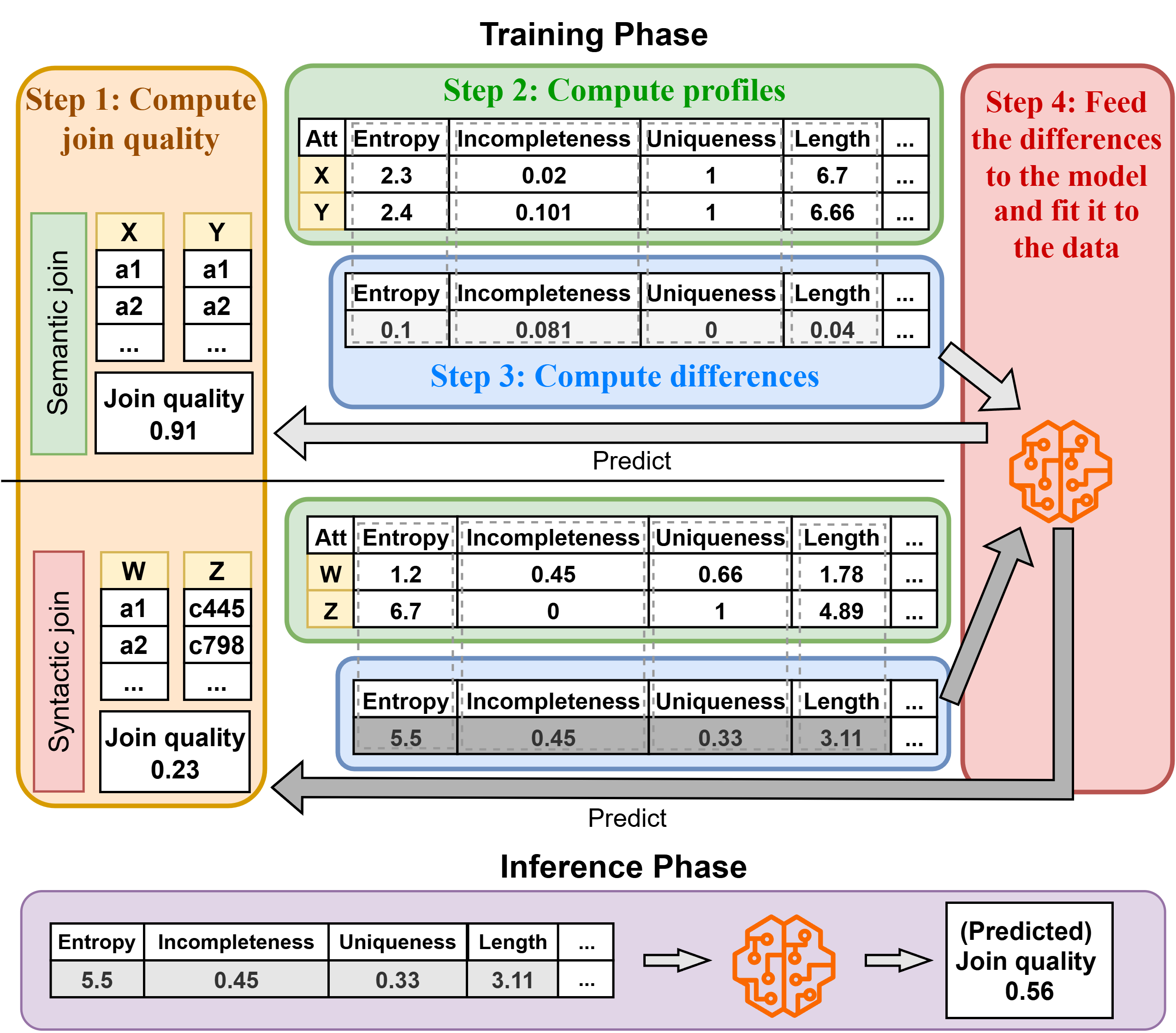}
		\caption{Training and inference phases of the predictive model}
		\label{fig:model_training}
	\end{center}
\end{figure}

Our core hypothesis is: \textit{\textbf{semantically related columns will have similar profiles}}. We assume that if two columns are semantically close, their underlying properties (e.g., incompleteness, entropy) will also be similar. Therefore, small pairwise differences between profile features indicate a high-quality join. The model is trained to predict the numeric joinability score from Section \ref{sec:join_quality_metric} using the differences between column profiles as input, as illustrated in Figure \ref{fig:model_training}.


During training, we first compute the true join quality for our ground truth, which serves as the prediction target. We then compute the profiles for each column pair and their absolute feature differences. These difference vectors are fed to the model, which adjusts its parameters to minimize the error between its predictions and the target values. The inference phase consists of computing the profile differences for a new column pair and feeding them to the trained model to generate a predicted quality value. To ensure generalizability and avoid costly fine-tuning, the model is trained only once on our diverse ground truth. This creates a general-purpose model applicable to new data lakes without retraining. Our goal is to create a \textbf{join quality predictive function}, $\mathcal{P}: (P(A),P(B)) \rightarrow \mathbb{R}_{[0,1]}$, that approximates the true quality metric $\mathcal{Q}$ with a much lower computational cost.

Our profiles are based on state-of-the-art techniques~\cite{DBLP:journals/vldb/AbedjanGN15}, using efficiently computed and highly descriptive features. These include general properties (e.g., uniqueness), value distributions (e.g., octiles), and syntactic properties (e.g., average string length), as detailed in Table \ref{tab:metadataList}. Profile comparison is done by computing the absolute difference for each feature pair. To handle features with different scales and prevent sensitivity to outliers, we apply Z-score normalization.

\begin{table*}
    \centering
    \caption{Summarized list of features composing a profile (bolded features are selected for the final model) \label{tab:metadataList}}
    \vspace{3mm}
    \resizebox{\textwidth}{!}{
        \begin{tabular}{|>{\centering\arraybackslash}m{3cm}|m{11.5cm}|>{\centering\arraybackslash}m{1.25cm}|}
            \hline
            \textbf{Category} & \centering\textbf{Metrics} & \textbf{Normalize}\\
            \hline
            General properties &
            \textbf{Cardinality}, \textbf{uniqueness}, incompleteness, \textbf{entropy} &
            Yes\\
            \hline
            Frequencies &
            Avg. frequency, \textbf{min. frequency}, \textbf{max frequency}, standard deviation of frequencies, \textbf{frequency IQR} &
            Yes\\
            \hline
            Octiles &
            Percentiles for octiles (12.5, 25, 37.5, etc.). \textbf{Selected: octiles 3, 4, and 6} (percentiles 37.5, 50, 75) &
            No\\
            \hline
            Frequency percentages &
            \textbf{Max. frequency percentage}, min. frequency percentage, \textbf{stdev of frequency percentages} &
            No\\
            \hline
            Frequent values &
            \textbf{Most frequent values}, most frequent values (Soundex) &
            No\\
            \hline
            Data type &
            Data type, specific data type, data type percentage, specific data type percentage &
            No\\
            \hline
            Value lengths &
            \textbf{Average string length}, \textbf{maximum string length}, \textbf{minimum string length} &
            Yes\\
            \hline
            Number of words &
            Number of words, \textbf{average number of words}, \textbf{minimum number of words}, \textbf{maximum number of words}, \textbf{standard deviation of number of words} &
            Yes\\
            \hline
            Specific values &
            \textbf{First value}, last value &
            Yes\\
            \hline
        \end{tabular}
    }
\end{table*}


\subsection{A generic regression model for joinability on data lakes}\label{sec:regression_model}

The predictive function $\mathcal{P}$ is implemented as a regression model. 
To maintain the scalability of our approach, we exclusively tested lightweight base regressors (e.g. LightGBM) to instantiate the model. Additionally, we conducted exhaustive feature selection tasks, as we expected several of the initial set of features to be redundant and/or irrelevant. Moreover, some of the features were costly to compute, so, unless their contribution was absolutely critical to the model, pruning them benefited the overall scalability of the approach.

\revision{Our feature selection pipeline used a multi-stage process combining filter, embedded, and wrapper methods to iteratively reduce the feature space with increasingly costly operations (the full workflow is available on the paper’s companion website). This process reduced the profile size from 65 individual metrics to 22 (indicated in Table \ref{tab:metadataList}). The final model achieved the lowest test errors across MSE, MedAE, and MAE metrics. This improvement likely stems from reduced overfitting through the elimination of irrelevant features.} 
The best performing model was implemented via Gradient Boosting, and underwent a fine-tuning process to obtain the best set of hyperparameters (while preventing excessive fine-tuning to the training data). The most relevant changes include a $learning\_rate = 0.05$, $subsample = 0.8$ and $min\_samples\_leaf = 10$. The test error exhibited by the model is less than 0.01, which represents an average relative error of less than 2\% of the original join quality score. 
The low prediction error strengthens our hypothesis that profile similarity effectively correlates with our join quality metric.

The resulting regression model enjoys a small feature space, which contributes to its generalization as we only keep the most relevant features that are likely to have a meaningful impact on predictive accuracy, regardless of the data lake. Developing a model capable of adapting to unseen contexts is key for our methodology, as we want to offer a flexible approach that does not require retraining or tuning to specific data, which would be costly and obtrusive. Hence, the model captures the general patterns of how profiles relate to join quality for our benchmark but leaves room for variability when executed in other data lakes. Moreover, as we do not rely on opaque structures such as embeddings, the join detection mechanism is fully explainable and transparent.

%% file: 5_experiments.tex
\section{Experimental evaluation}\label{sec:experiments}

This section experimentally compares our approach with state-of-the-art data discovery systems.

\begin{figure*}
	\centering
	\includegraphics[scale=0.6]{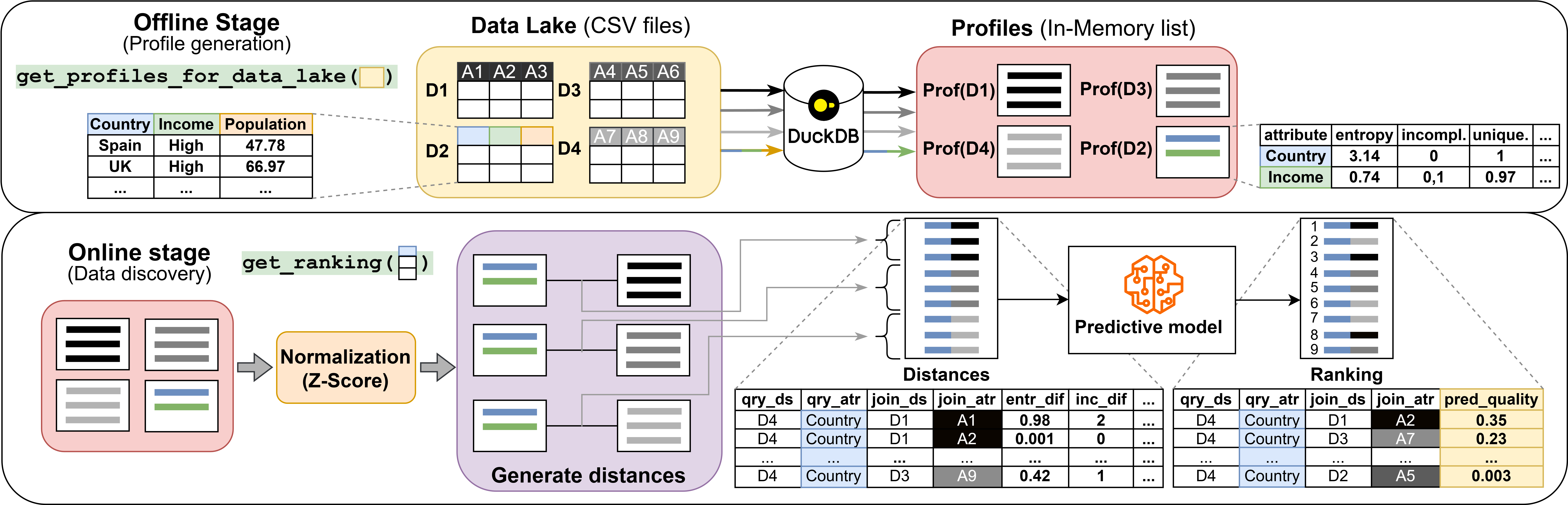}
	\caption{\sys~system architecture}\label{fig:pipeline}
\end{figure*}

\stitle{System description} 
Our approach is implemented in a Python system called \sys~\cite{DBLP:conf/edbt/MaynouN25}\footnote{\revision{\sys~is openly available here: \href{https://github.com/dtim-upc/FREYJA}{https://github.com/dtim-upc/FREYJA}}}, which uses DuckDB to efficiently compute column profiles. Figure \ref{fig:pipeline} depicts its architecture. The offline stage consists solely of computing data lake profiles, which is the only preprocessing required. This minimal overhead allows new data to be added without costly model retraining. The online stage implements the pipeline from Section \ref{sec:predictingQ}: after normalization, distances are computed between the query column and all others, and a final ranking is generated based on the model-inferred join quality. \revision{\sys's repository provides implementation details and execution instructions. The companion website contains extensive explanations of extra experiments and access to supplemental material.}

\stitle{Experimental setup and baselines} 
All experiments were run on a machine with an Intel Core i7-10700K CPU, 32GB RAM, an Nvidia GTX 3070 GPU (8GB VRAM), and a 512GB SSD. \revision{We compare \sys~against six state-of-the-art systems: D3L, Starmie, SANTOS, DeepJoin, WarpGate, and KGLiDS. Starmie, DeepJoin, and WarpGate are embedding-based (the first two require model training). D3L and KGLiDS use a hybrid approach (TRL with profiles), while SANTOS uses a knowledge base. As DeepJoin and WarpGate are not open-source, we used our own implementations (available in \sys's repository). We also include an ablation study using a model based solely on $\mathcal{J}$ to highlight the contribution of $K$.}

\stitle{Benchmarks} 
Join discovery benchmarks consist of a data lake (tabular datasets) and a ground truth listing relevant joins for a set of query columns. To evaluate a system, we measure the average number of relevant joins found across all queries. We employ eight widely-used benchmarks: SANTOS (Small, Large)~\cite{khatiwada2023santos}, TUS (Small, Large)~\cite{nargesian2018table}, D3L$_\text{BM}$~\cite{BogatuFP020}, \revision{OmniMatch (OM$_\text{CG}$, OM$_\text{CR}$)~\cite{DBLP:journals/pvldb/KoutrasZQLIFKK25}}, and our own \sys$_{\text{BM}}$. Table \ref{tab:benchmarks} summarizes their characteristics, distinguishing between synthetic and non-synthetic data. D3L$_\text{BM}$ and \sys$_{\text{BM}}$ are manually annotated, while the others are generated via dataset partitioning. In synthetic benchmarks the ground truth is obtained in an automated fashion. For benchmarks with many queries (D3L$_\text{BM}$, TUS Large, \revision{OM$_\text{CG}$, OM$_\text{CR}$}), we randomly sampled a subset for our experiments. For our benchmark, we selected 50 query columns with at least 10 semantic joins each. The SANTOS Large benchmark, which lacks a ground truth, is used only for efficiency tests. \revision{All ground truths are available in \sys's repository.}

\begin{table}
    \caption{Characteristics of the considered benchmarks.}
    \vspace{3mm}
    \centering
    \Large
    \resizebox{\columnwidth}{!} {
        \begin{tabular}{ccccccc}
            \toprule
            \textbf{Benchmark} & \textbf{Type} & \textbf{Size (GB)} & \textbf{Tables} & \textbf{Queries} & \textbf{Columns} & \textbf{Avg. \#rows} \\
            \midrule
            \sys$_{\text{BM}}$ & NS & 0.13 & 159 & 50 & 1,316 & 12,613 \\
            SANTOS Small & S & 0.42 & 550 & 50 & 6,322 & 6,921  \\
            TUS Small & S & 1.12 & 1,530 & 125 & 14,810 & 4,453 \\
            D3L$_\text{BM}$ & NS & 1.31 & 654 & 100 & 8,781 & 16,926  \\
            TUS Large & S & 1.38 & 5,044 & 100 & 54,590 & 1,910 \\
            \revision{OM${_\text{CG}}$} &  \revision{S} &  \revision{0.09} &  \revision{550} &  \revision{50} &  \revision{2,641} &  \revision{1,453} \\
            \revision{OM${_\text{CR}}$} &  \revision{S} &  \revision{0.01} &  \revision{600} &  \revision{50} &  \revision{2,657} &  \revision{209} \\
            \midrule
            SANTOS Large & S & 10.9 & 11,086 & 80 & 121,498 & 7,718\\
            \bottomrule
        \end{tabular}
    }
    \\
    \vspace{0.1cm}
    \footnotesize
    S = Synthetic data lake, NS = Non-synthetic data lake 
    \label{tab:benchmarks}
\end{table}

\stitle{Metrics} 
Effectiveness is measured with Precision at $k$ ($P@k$) and Recall at $k$ ($R@k$), standard metrics for this task~\cite{khatiwada2023santos, nargesian2018table, fan2022semantics, BogatuFP020}. For each query, we evaluate the top-$k$ ranked candidates. $P@k$ is the fraction of relevant joins in the top-$k$ list, while $R@k$ is the fraction of all possible relevant joins found in the top-$k$. Since the total number of relevant joins varies per query, we normalize $R@k$ by the maximum possible recall for each query. The value of $k$ varies by benchmark: 10 for SANTOS Small and \sys$_{\text{BM}}$, \revision{30 for OM}, 60 for TUS, and 100 for D3L$_\text{BM}$. Efficiency is measured by \textit{preparation time} (to process the data lake and build necessary structures) and \textit{query time} (to generate a ranking). For \sys, preparation is profile computation, and querying involves computing profile distances and model inference. We also measure the memory footprint of the generated structures.

\renewcommand{\arraystretch}{1.1}

\begin{table*}
    \caption{Execution times for each tool and benchmark} 
    \vspace{3mm}
    \centering
    \resizebox{\textwidth}{!} {
        \begin{tabular}{ccccccccc}
            \toprule
            \textbf{Benchmark} & \textbf{Process} & \textbf{SANTOS} & \textbf{Starmie} & \textbf{D3L} & \textbf{DeepJoin} & \textbf{KGLiDS} & \textbf{WarpGate} & \textbf{\sys} \\
            \midrule
            \multirow{2}{*}{\sys$_{\text{BM}}$} & Prep. & 1 h, 40 m & 34 m & 6 m, 14 s & 42 m, 33 s & 28 m, 34 s & 14 m, 35 s & \textbf{7 s} \\ 
                                   & Query & \revision{245 $\pm$ 2.99 s} & \revision{0.32 $\pm$ 0.02 s} & \revision{130 $\pm$ 2.56 s} & \revision{0.02 $\pm$ 0.0 s} & \revision{0.07 $\pm$ 0.01 s} & \textbf{\revision{0.01 $\pm$ 0.00 s}} & \textbf{\revision{0.01 $\pm$ 0.00 s}} \\ 
            \multirow{2}{*}{SANTOS Small} & Prep. & 2 h, 51 m & 1 h, 40 m & 15 m, 25 s & 1 h, 23 m & 3 h, 38 m & 35 m, 44 s & \textbf{48 s} \\ 
                                   & Query & \revision{255 $\pm$ 2.99 s} & \revision{1.82 $\pm$ 0.06 s} & \revision{142 $\pm$ 2.48 s} & \revision{0.2 $\pm$ 0.01 s} & \revision{0.18 $\pm$ 0.02 s} & \revision{0.08 $\pm$ 0.03 s} & \textbf{\revision{0.07 $\pm$ 0.03 s}} \\ 
            \multirow{2}{*}{TUS Small} & Prep. & 5 h & 6 h, 54 m & 57 m, 37 s & 3 h, 59 m & 13 h, 50 m & 1 h, 38 s & \textbf{1 m, 59 s} \\ 
                                   & Query & \revision{249 $\pm$ 2.96 s} & \revision{2.56 $\pm$ 0.06 s} & \revision{173 $\pm$ 2.46 s} & \revision{0.36 $\pm$ 0.02 s} & \revision{0.34 $\pm$ 0.03 s} & \revision{0.19 $\pm$ 0.01 s} & \textbf{\revision{0.17 $\pm$ 0.04 s}}\\ 
            \multirow{2}{*}{D3L$_\text{BM}$} & Prep. & 4 h, 6 m & 2 h, 34 m & 49 m, 7 s & 4 h, 32 m & 8 h, 15 m & 1 h, 20 m & \textbf{1 m, 11 s} \\ 
                                   & Query & \revision{267 $\pm$ 3.08 s} & \revision{3.63 $\pm$ 0.06 s} & \revision{153 $\pm$ 2.46 s} & \revision{0.49 $\pm$ 0.03 s} & \revision{0.27 $\pm$ 0.02 s} & \revision{0.22 $\pm$ 0.01 s} & \textbf{\revision{0.15 $\pm$ 0.03 s}} \\ 
            \multirow{2}{*}{TUS Large} & Prep. & 10 h, 23 m & 15 h, 34 m & 1 h, 15 m & 20 h, 35 m & 44 h, 30 m & 2 h, 55 m & \textbf{6 m, 22 s}\\ 
                                   & Query & \revision{275 $\pm$ 3 s} & \revision{4.91 $\pm$ 0.08 s} & \revision{195 $\pm$ 2.77 s} & \revision{2.6 $\pm$ 0.03 s} & \revision{1.1 $\pm$ 0.03 s} & \revision{0.57 $\pm$ 0.04 s} & \textbf{\revision{0.55 $\pm$ 0.02 s}}\\ 
            \multirow{2}{*}{SANTOS Large} & Prep. & 18 h, 24 m & 38 h, 32 m & 6 h, 35 m & 74 h, 22 m & 147 h, 35 m & 13 h, 19 m & \textbf{16 m, 22 s}\\ 
                                   & Query & \revision{287 $\pm$ 3.08 s} & \revision{19.25 $\pm$ 0.15 s} & \revision{243 $\pm$ 2.84 s} & \revision{3.77 $\pm$ 0.04 s} & \revision{2.94 $\pm$ 0.04 s} & \revision{1.13 $\pm$ 0.03 s} & \textbf{\revision{0.87 $\pm$ 0.04 s}}\\ 
            \multirow{2}{*}{\revision{OM${_\text{CG}}$}} & Prep. & \revision{1 h, 33 m} & \revision{17 m, 47 s} & \revision{3 m, 40 s} & \revision{8 m, 30 s} & \revision{1 h, 20 m} & \revision{16 m, 58 s} & \textbf{\revision{12 s}}\\
                                   & Query & \revision{238 $\pm$ 2.73 s} & \revision{0.09 $\pm$ 0.02 s} & \revision{112 $\pm$ 1.68 s} & \revision{0.02 $\pm$ 0.0 s} & \revision{0.1 $\pm$ 0.01 s} & \textbf{\revision{0.01 $\pm$ 0.00 s}} & \textbf{\revision{0.01 $\pm$ 0.00 s}}\\
            \multirow{2}{*}{\revision{OM${_\text{CR}}$}} & Prep. & \revision{1 h, 24 m} & \revision{15 m, 22 s} & \revision{3 m, 32 s} & \revision{5 m, 49 s} & \revision{1 h, 23 m} & \revision{10 m, 30 s} & \textbf{\revision{23 s}}\\
                                   & Query & \revision{239 $\pm$ 2.73 s} & \revision{0.10 $\pm$ 0.02 s} & \revision{110 $\pm$ 1.68 s} & \revision{0.02 $\pm$ 0.0 s} & \revision{0.1 $\pm$ 0.01 s} & \textbf{\revision{0.01 $\pm$ 0.00 s}} & \textbf{\revision{0.01 $\pm$ 0.00 s}}\\ 
        \bottomrule
        \end{tabular}
    }
    \label{tab:efficiency}
\end{table*}

\subsection{Efficiency}\label{sec:exp_efficiency} 

\sys's primary advantage over state-of-the-art systems lies in its high efficiency, particularly during the preparation phase. As shown in Table \ref{tab:efficiency}, \sys~outperforms competitors by orders of magnitude in preprocessing costs (detailed in Table \ref{tab:preprocessing_stages}). \sys~also achieves faster query times by vectorizing distance computations, although these gains are less pronounced due to the already high efficiency of LSH-based indexes (e.g. WaprGate, DeepJoin). \sys's preprocessing only requires computing profiles, which is highly efficient. In contrast, baselines often create and manipulate high-dimensional embeddings, incurring high execution costs. These TRL-based approaches require GPUs and large RAM, making them impractical on modest hardware. \sys~imposes no specific hardware demands; it can run on less powerful systems, albeit slower. Since profile computation is an independent task, our approach is highly scalable and parallelizable.


\revision{\stitle{Preprocessing stages} To guarantee a fair comparison across preprocessing times, we have separated the preprocessing costs for the SANTOS Big benchmark (the biggest of the data lakes considered) into several stages (Table \ref{tab:preprocessing_stages}). This ensures a fairer comparison across systems, as it is clear how much time is spent on each specific task:
\begin{itemize}[leftmargin=*]
    \item \textit{Model fine-tuning}: both Starmie and DeepJoin require fine-tuning a model. The former trains a column encoder via contrastive learning, whereas the latter fine-tunes a model with new training data. Building this training data incurs high computational costs, as it requires finding positive examples for the model, that is, columns from the benchmark with high similarity. The larger and more heterogeneous the benchmark, the harder to do so.
    \item \textit{Semantic inference}: on one hand, KGLiDS richly annotates data by building a knowledge graph capturing the semantics of datasets and code pipelines. On the other hand, SANTOS labels columns with types from YAGO and discovers semantic relationships (both via YAGO and a “synthesized” KB from the lake).
    \item \textit{Generate representations}: all systems, except \sys~and SANTOS, employ embedding models (optionally fine-tuned) to generate embedding representations of columns. This operation tends to be costly, specially as the complexity of the embeddings increases and in scenarios where dedicated GPUs are not employed. To complement the embeddings, KGLiDS generate profiles (besides the KB previously indicated) and D3L extract syntactic features (name, format, values and distributions). SANTOS relies solely on their KB knowledge. Finally, \sys~just needs to compute lightweight column profiles.
\end{itemize}
\sys~does not need to either fine-tune a model or perform semantic annotation, and we avoid the use of embeddings and focus solely on profiles. Hence, the obtention of the structures to represent the columns is done substantially faster as we prevent executing very costly operations to manipulate high-dimension artifacts. \sys's speed-up w.r.t. to the baselines ranges from 24x to 541x, so our preprocessing costs are orders of magnitude faster.}


\begin{table}
    \caption{\revision{Preprocessing times divided by stages (Santos Big benchmark)}} 
    \centering
    \vspace{0.2cm}
    \large{\resizebox{\linewidth}{!} {
        \begin{tabular}{cccccc}
        \toprule
        \revision{\textbf{System}} & \revision{\makecell{\textbf{Model} \\ \textbf{tuning}}} & \revision{\makecell{\textbf{Semantic} \\ \textbf{inference}}} & \revision{\makecell{\textbf{Generate} \\ \textbf{representations}}} & \revision{\textbf{Total}} & \revision{\makecell{\sys's \\ \textbf{Speed-up}}} \\
        \midrule
        \revision{KGLiDS$^*$$^\ddagger$$^\dagger$} & \revision{-} & \revision{4 h, 32 m} & \revision{142 h, 53 m} & \revision{147 h, 35 m} & \revision{x541}\\
        \revision{DeepJoin$^*$} & \revision{71 h, 47 m} & \revision{-} & \revision{2 h, 35 m} & \revision{74 h, 22 m} & \revision{x272}\\
        \revision{Starmie$^*$} & \revision{34 h, 48 m} & \revision{-} & \revision{3 h, 44 m} & \revision{38 h, 32 m} & \revision{x141}\\
        \revision{SANTOS$^\ddagger$} & \revision{-} & \revision{18 h, 24 m} & \revision{-} & \revision{18 h, 24 m} & \revision{x67}\\
        \revision{WarpGate$^*$} & \revision{-} & \revision{-} & \revision{13 h, 19 m} & \revision{13 h, 19 m} & \revision{x49}\\
        \revision{D3L$^*$$^\dagger$} & \revision{-} & \revision{-} & \revision{6 h, 35 m} & \revision{6 h, 35 m} & \revision{x24}\\
        \revision{\sys$^\dagger$} & \revision{-} & \revision{-} & \revision{16 m, 22 s} & \revision{16 m, 22 s} & \revision{x1}\\
        \bottomrule
        \end{tabular}
    }}
    \\
    \vspace{0.1cm}
    \footnotesize
    \revision{$^*$Embedding-based \quad $^\ddagger$Knowledge-based \quad $^\dagger$Based on descriptive statistics}
    \label{tab:preprocessing_stages}
\end{table}


\begin{table*}
    \caption{Memory footprints of the selected systems (in MB) and the percentage they represent over the benchmark size} 
    \vspace{3mm}
    \centering
    \resizebox{\textwidth}{!} {
        \begin{tabular}{cccccccc}
        \toprule
        \textbf{Benchmark} & \textbf{SANTOS} & \textbf{Starmie} & \textbf{D3L} & \textbf{DeepJoin} & \textbf{KGLiDS} & \textbf{WarpGate} & \textbf{\sys} \\
        \midrule
        \sys$_{\text{BM}}$ & 1,640 (1,242\%) & 494 (374\%) & 23.8 (18\%) & 154 (118\%) & 15.36 (11.53\%) & 10.4 (8\%) & \textbf{0.37 (0.28\%)}\\
        SANTOS Small & 2,084 (495\%) & 519 (123\%) & 82.7 (19.64\%) & 158 (37\%) & 110 (26.19\%) & 53.9 (12.83\%) & \textbf{2.01 (0.47\%)}\\
        TUS Small & 2,367 (217\%) & 547 (50.18\%) & 203 (18.62\%) & 524 (46.78\%) & 1160 (103\%) & 119 (10.64\%) & \textbf{8.07 (0.74\%)}\\
        D3L$_\text{BM}$ & 2,245 (176\%) & 528 (42.58\%) & 104 (8.19\%) & 214 (16.33\%) & 244 (18.63\%) & 70.5 (5.38\%) & \textbf{3.83 (0.29\%)}\\
        TUS Large & 3,546 (262\%) & 656 (48.59\%) & 682 (50.52\%) & 591 (42.82\%) & 2083 (151\%) & 438 (31.76\%) & \textbf{24.47 (1.80\%)}\\
        SANTOS Large & 5,574 (52.59\%) & 848 (8.00\%) & 987 (9.31\%) & 1,540 (14.12\%) & 22,340 (205\%) & 822 (7.54\%) & \textbf{40.94 (0.38\%)}\\
        \revision{OM${_\text{CG}}$} & \revision{2,289 (1,242\%)} & \revision{499 (554\%)} & \revision{26.3 (29.2\%)} & \revision{143 (159\%)} & \revision{196.6 (218\%)} & \revision{21.1 (23.4\%)} & \revision{\textbf{1.27 (1.41\%)}}\\
        \revision{OM${_\text{CR}}$} & \revision{1,967 (19,670\%)} & \revision{497 (4970\%)} & \revision{26.5 (265\%)} & \revision{102 (1,020\%)} & \revision{216.2 (2,162\%)} & \revision{21.3 (210\%)} & \revision{\textbf{1.35 (13.5}\%)}\\
        \bottomrule
        \end{tabular}
    }
    \label{tab:memory}
\end{table*}

\stitle{Analysis of the memory footprint} To complement the previous analysis we can examine the size of the structures created by each system. This is a relevant factor, as commonly-employed artifacts, such as embeddings or knowledge graphs, require significant storage space, further increasing the hardware requirements of approaches that leverage these components. This refers both to in-memory (RAM) loading as well as permanent storage (HDDs/SSDs). Table \ref{tab:memory} showcases the space occupied by each system to preprocess each benchmark, accompanied by its respective memory overhead with regards to the total size of the benchmark. We can observe how \sys~presents, by a large margin, the smallest cost in all scenarios, as in most cases the combined size of the profiles is less than 1\% of the memory occupied by the respective benchmark. Again, we can take the SANTOS Large benchmark to further highlight the differences, as \sys's profiles occupy between 20x-540x less memory space than the structures obtained by other approaches 
and a total of 0.38\% of the original size of the benchmark. Hence, \sys's approach is easily adaptable to contexts with limited storage capacities, as processing large quantities of data does not imply equally large memory demands and the created artifacts will always represent a negligible fraction of the data lake's size.

\begin{figure}
    \centering
    \includegraphics[width=\columnwidth]{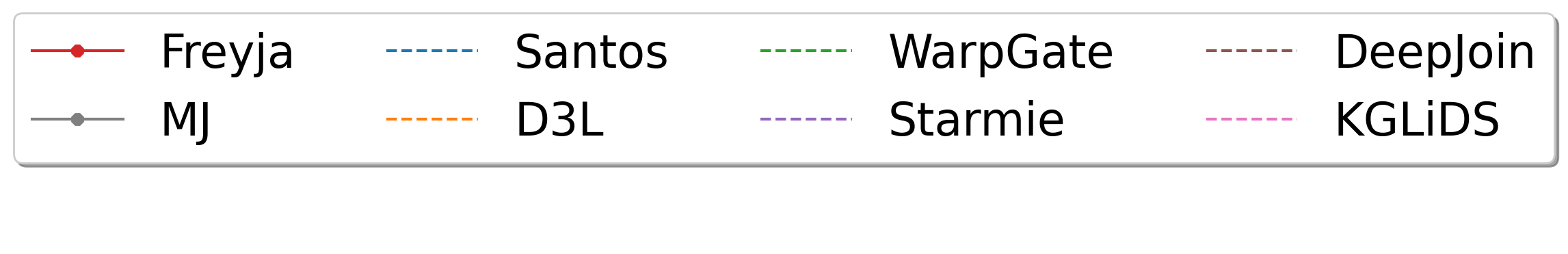}
    
    \begin{subfigure}{0.49\columnwidth}
        \centering
        \caption{Preparation time}
        \includegraphics[width=0.99\textwidth]{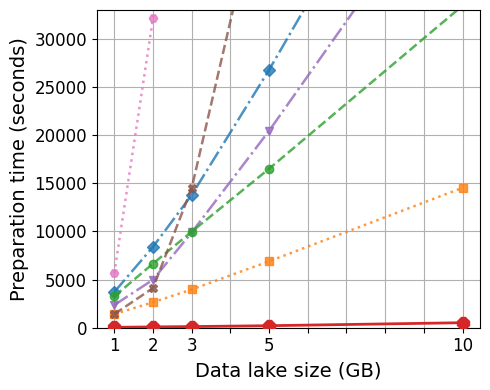}
        \label{fig:scalability_pre}
    \end{subfigure}
    \hfill
    \begin{subfigure}{0.49\columnwidth}
        \centering
        \caption{Query time}
        \includegraphics[width=0.99\textwidth]{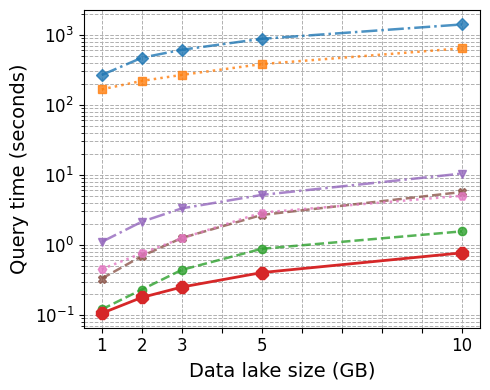}
        \label{fig:scalability_query}
    \end{subfigure}

    \caption{\revision{Scalability on synthetic benchmarks}}
    \label{fig:scalability}
\end{figure}

\stitle{Scalability analysis} To systematically showcase the scalability of \sys, we built a different set of synthetic data lakes where we could normalize the size of its datasets. That is, differently from the benchmarks previously used, each file (i.e. dataset) has the same size, which guarantees a more stable comparison and a better showcasing of the capacity to scale to increasing quantities of data. More precisely, we developed five benchmarks of 1 to 10 GB, each composed of a varying number of files, all of them with a weight of 1MB. Figure \ref{fig:scalability} presents the scalability analysis for progressively larger data lakes, both for preprocessing and query time (the latter in logarithmic scale).
Figure \ref{fig:scalability_pre} clearly displays that \sys~outperforms all alternatives in preprocessing time, as it scales in a completely linear fashion given that profiles are computed independently and demand the same computation requirements regardless of the size of the underlying set of values. Most of the other tools also scale linearly, but a slight elbow shape can be seen starting from the 3GB data lake, which evidences that these systems are already experiencing overhead from data processing for relatively small data lakes. Figure \ref{fig:scalability_query} showcases that \sys's lightweight approach is faster than the index-based similarity searches of the other methods, even for increasingly large benchmarks.

    

\subsection{Effectiveness}\label{sec:exp_effectiveness}

\begin{figure}[t!]
    \centering
    \includegraphics[width=\columnwidth]{img/effectiveness/p_r_legend_extended.png}
    
    \textbf{Synthetic data lakes}
    \vspace{0.2cm}
    
    \begin{subfigure}{0.49\columnwidth}
        \centering
        \caption{$P@k$ on SANTOS Small}
        \includegraphics[width=0.97\textwidth]{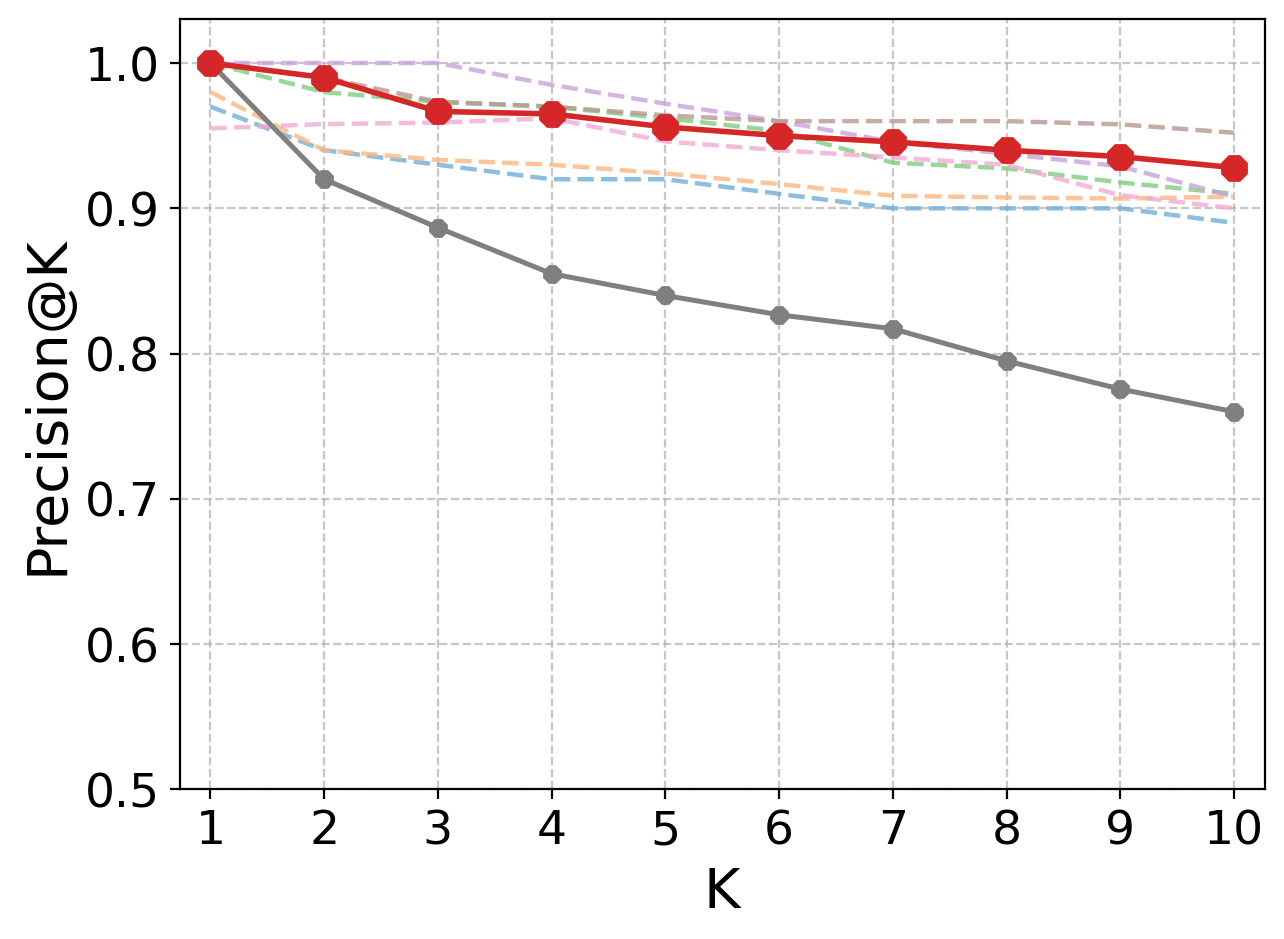}
        \label{fig:santos_small_p}
    \end{subfigure}
    \hfill
    \begin{subfigure}{0.49\columnwidth}
        \centering
        \caption{$R@k$ on SANTOS Small}
        \includegraphics[width=0.97\textwidth]{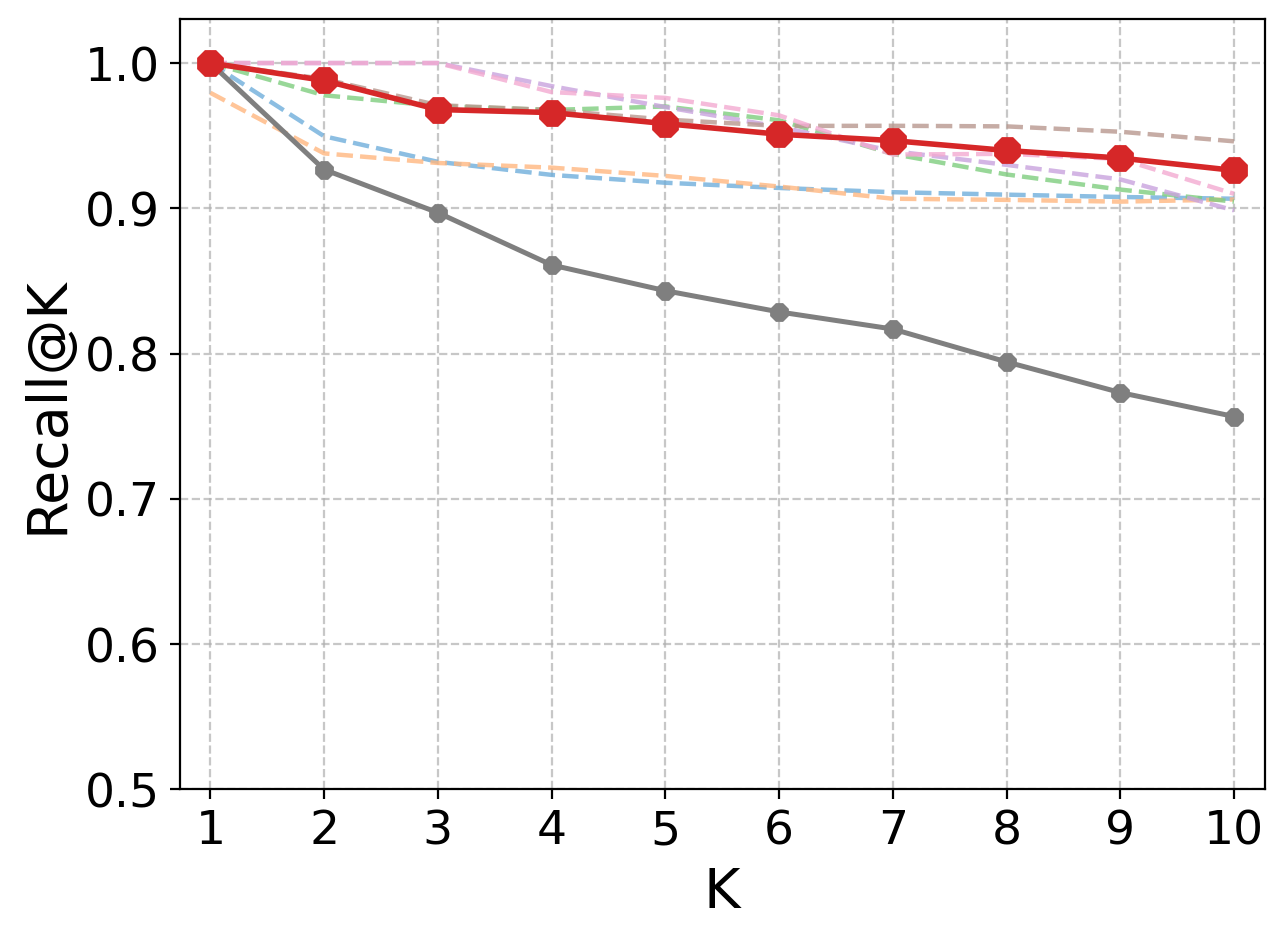}
        \label{fig:santos_small_r}
    \end{subfigure}

    \begin{subfigure}{0.49\columnwidth}
        \centering
        \caption{$P@k$ on TUS Small}
        \includegraphics[width=0.97\textwidth]{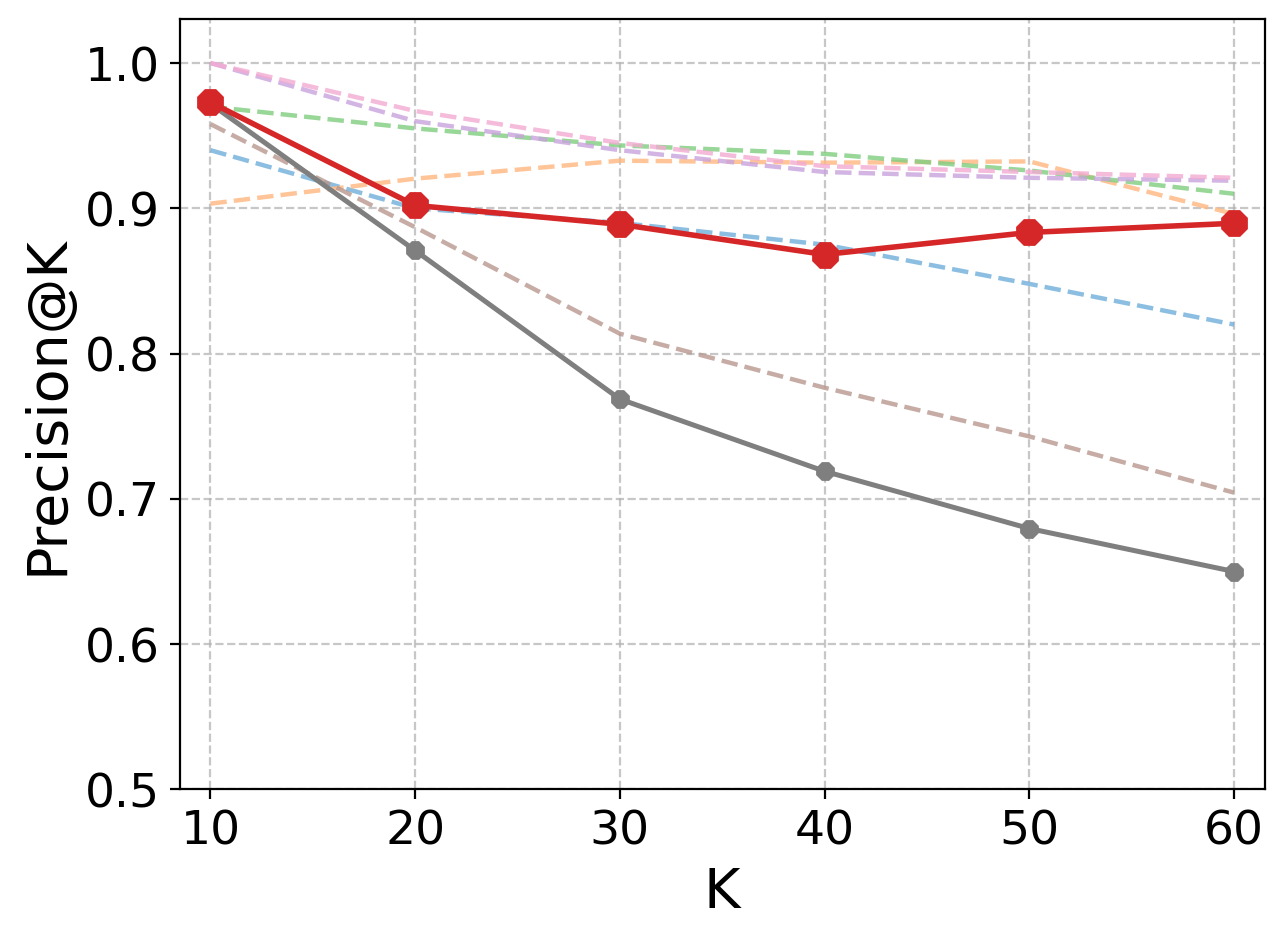}
        \label{fig:tus_small_p}
    \end{subfigure}
    \hfill
    \begin{subfigure}{0.49\columnwidth}
        \centering
        \caption{$R@k$ on TUS Small}
        \includegraphics[width=0.97\textwidth]{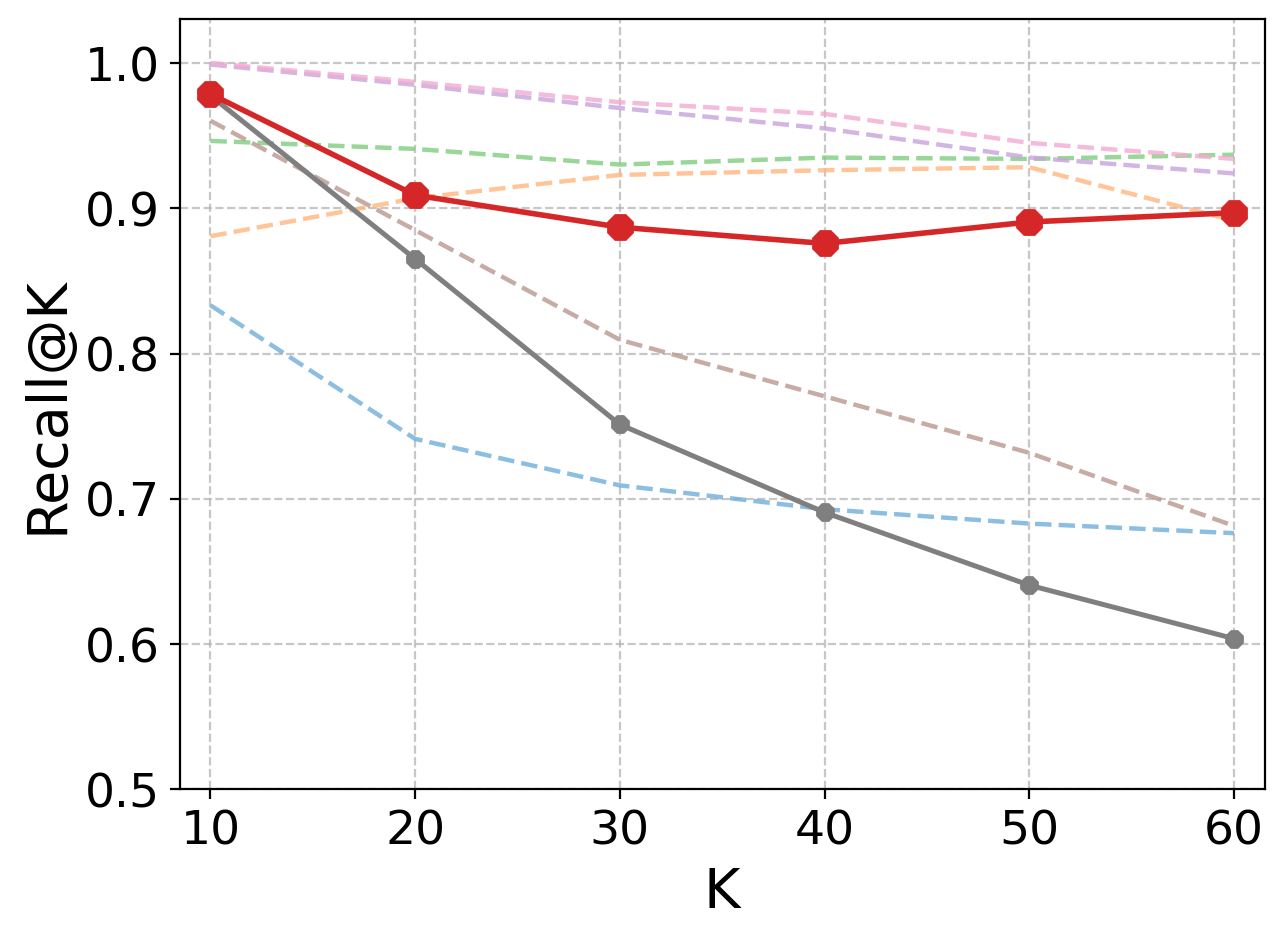}
        \label{fig:tus_small_r}
    \end{subfigure}

    \begin{subfigure}{0.49\columnwidth}
        \centering
        \caption{$P@k$ on TUS Large}
        \includegraphics[width=0.97\textwidth]{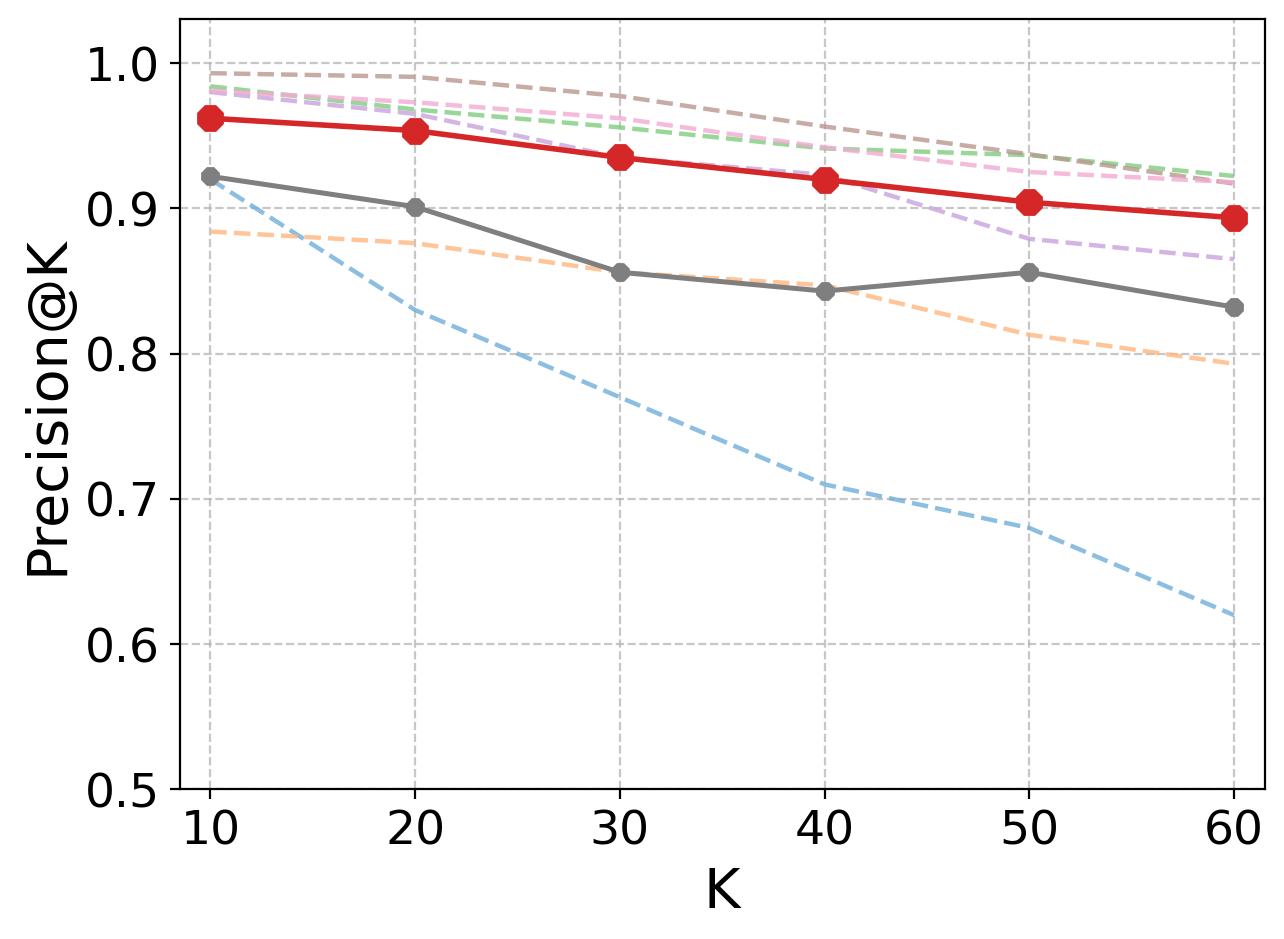}
        \label{fig:tus_big_p}
    \end{subfigure}
    \hfill
    \begin{subfigure}{0.49\columnwidth}
        \centering
        \caption{$R@k$ on TUS Large}
        \includegraphics[width=0.97\textwidth]{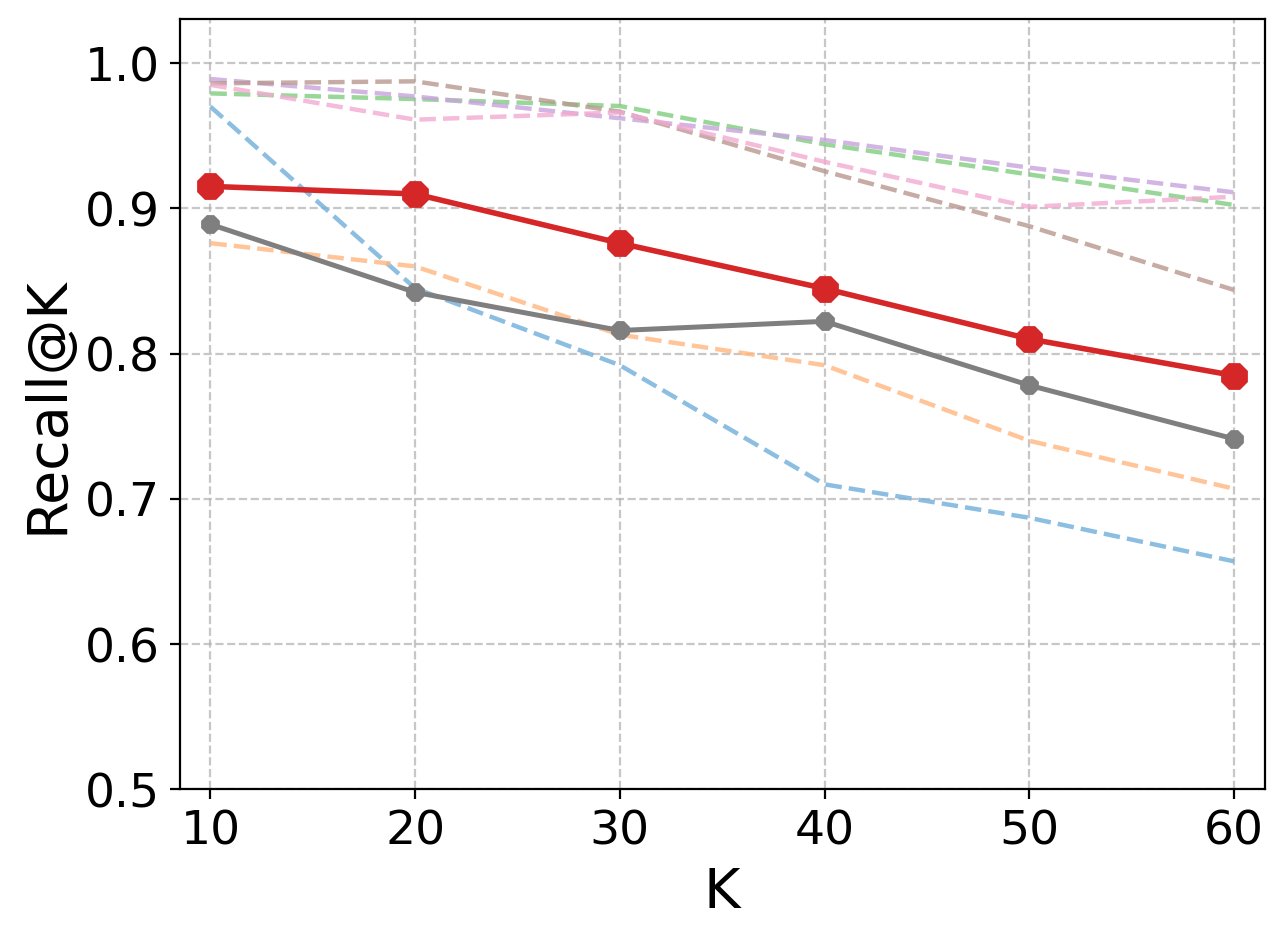}
        \label{fig:tus_big_r}
    \end{subfigure}

    \begin{subfigure}{0.49\columnwidth}
        \centering
        \caption{$P@k$ on OM${_\text{CG}}$}
        \includegraphics[width=0.97\textwidth]{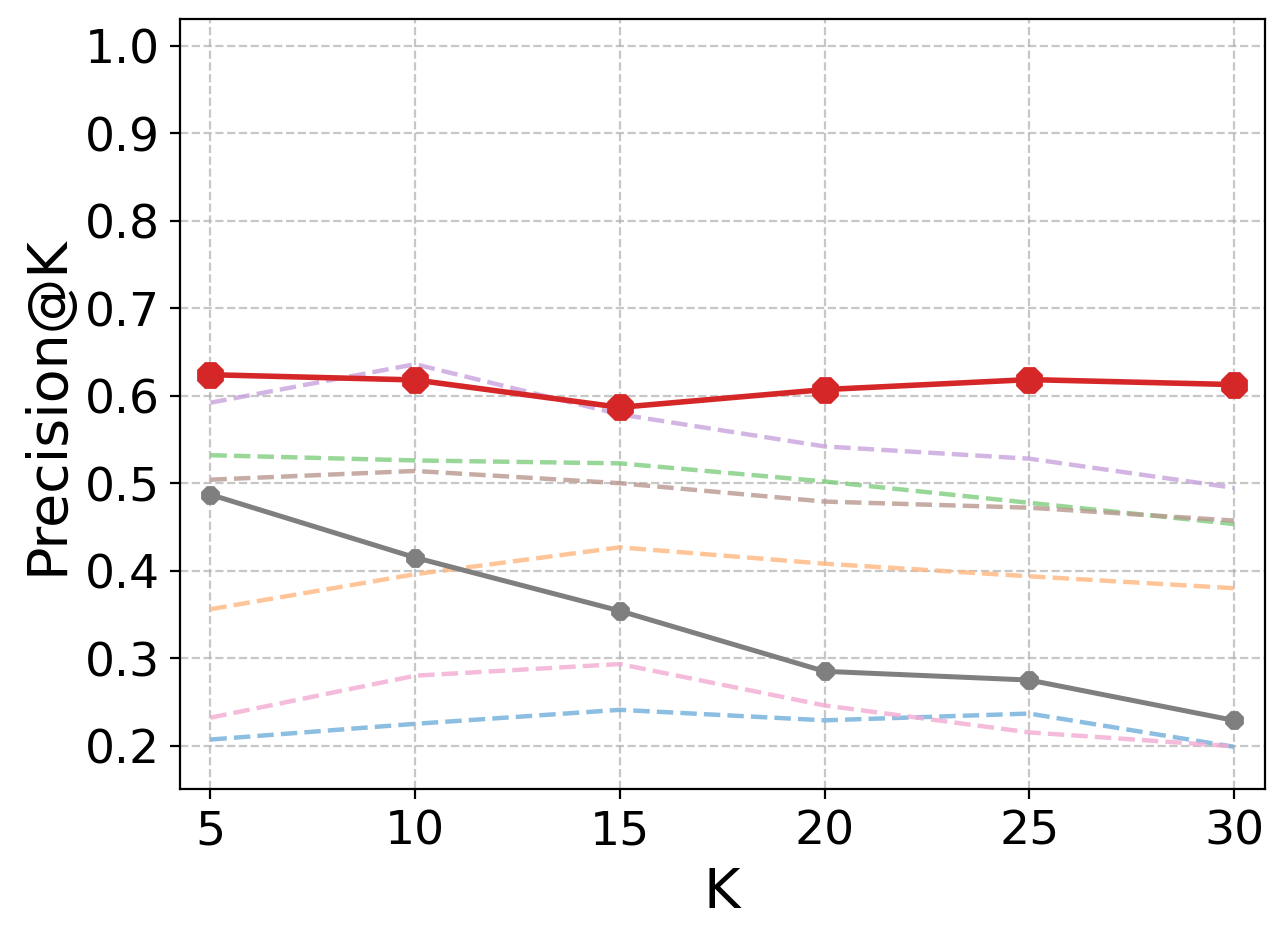}
        \label{fig:om_cg_p}
    \end{subfigure}
    \hfill
    \begin{subfigure}{0.49\columnwidth}
        \centering
        \caption{$R@k$ on OM${_\text{CG}}$}
        \includegraphics[width=0.97\textwidth]{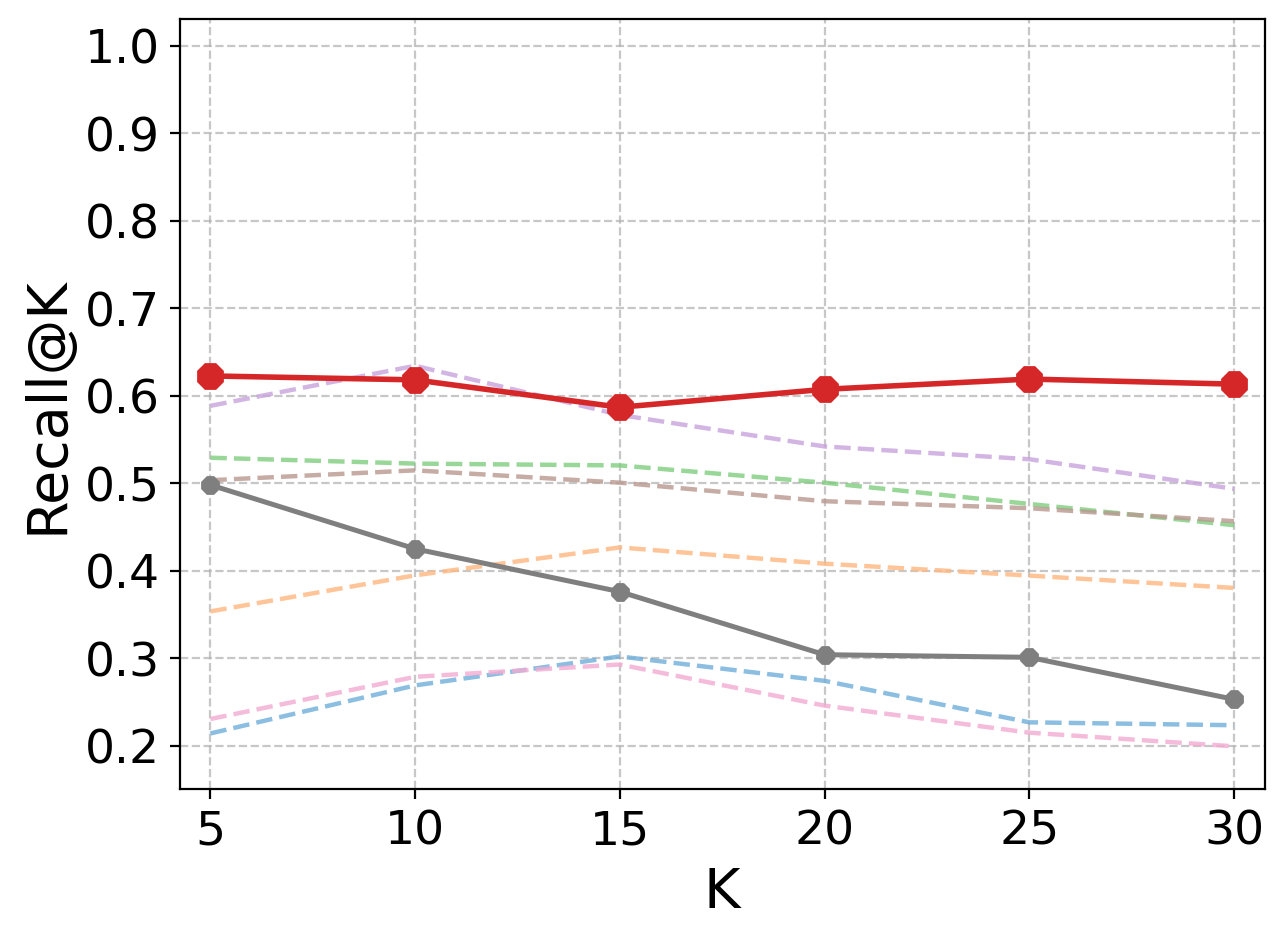}
        \label{fig:om_cg_r}
    \end{subfigure}

    \begin{subfigure}{0.49\columnwidth}
        \centering
        \caption{$P@k$ on OM${_\text{CR}}$}
        \includegraphics[width=0.97\textwidth]{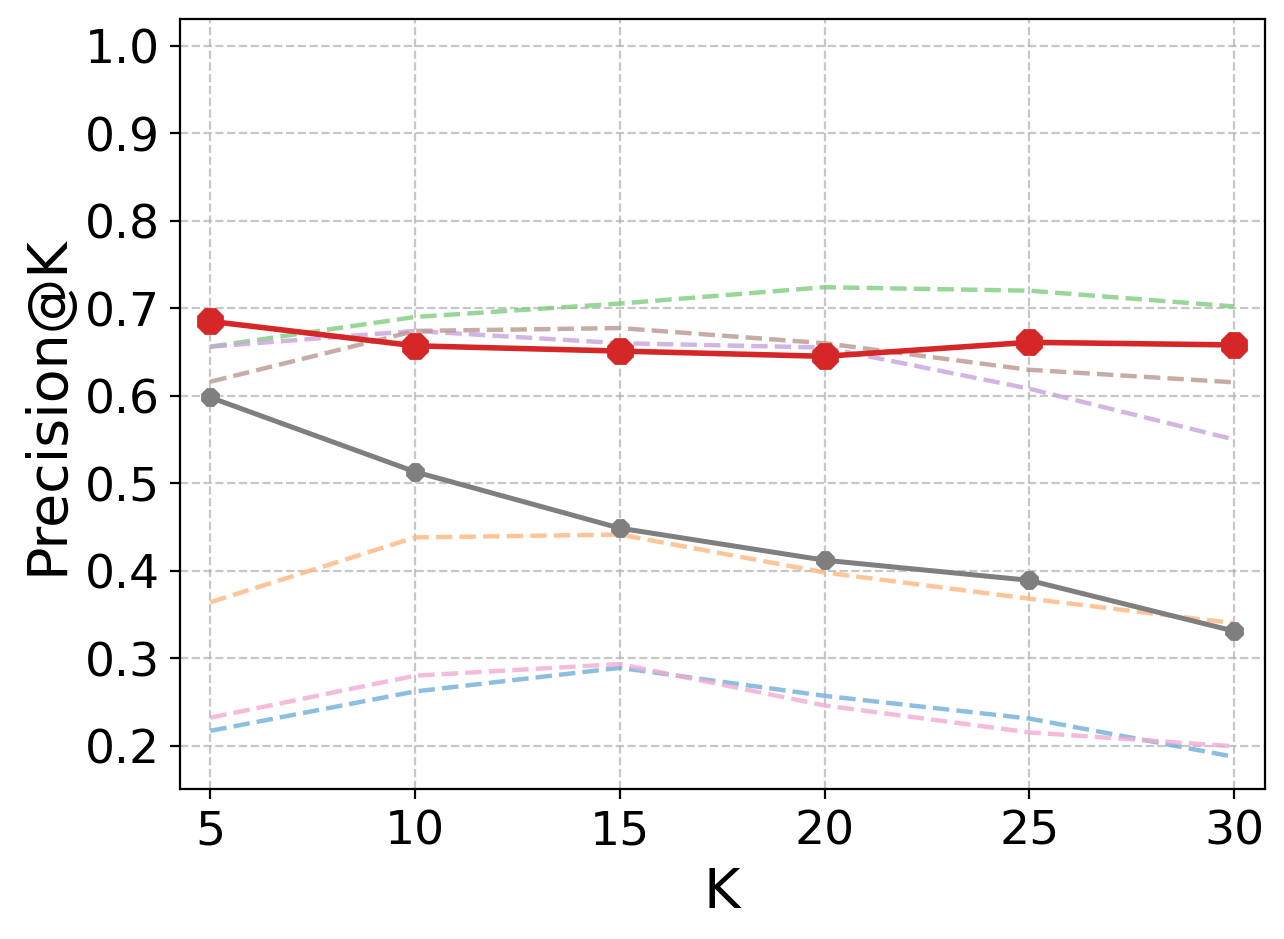}
        \label{fig:om_cr_p}
    \end{subfigure}
    \hfill
    \begin{subfigure}{0.49\columnwidth}
        \centering
        \caption{$R@k$ on OM${_\text{CR}}$}
        \includegraphics[width=0.97\textwidth]{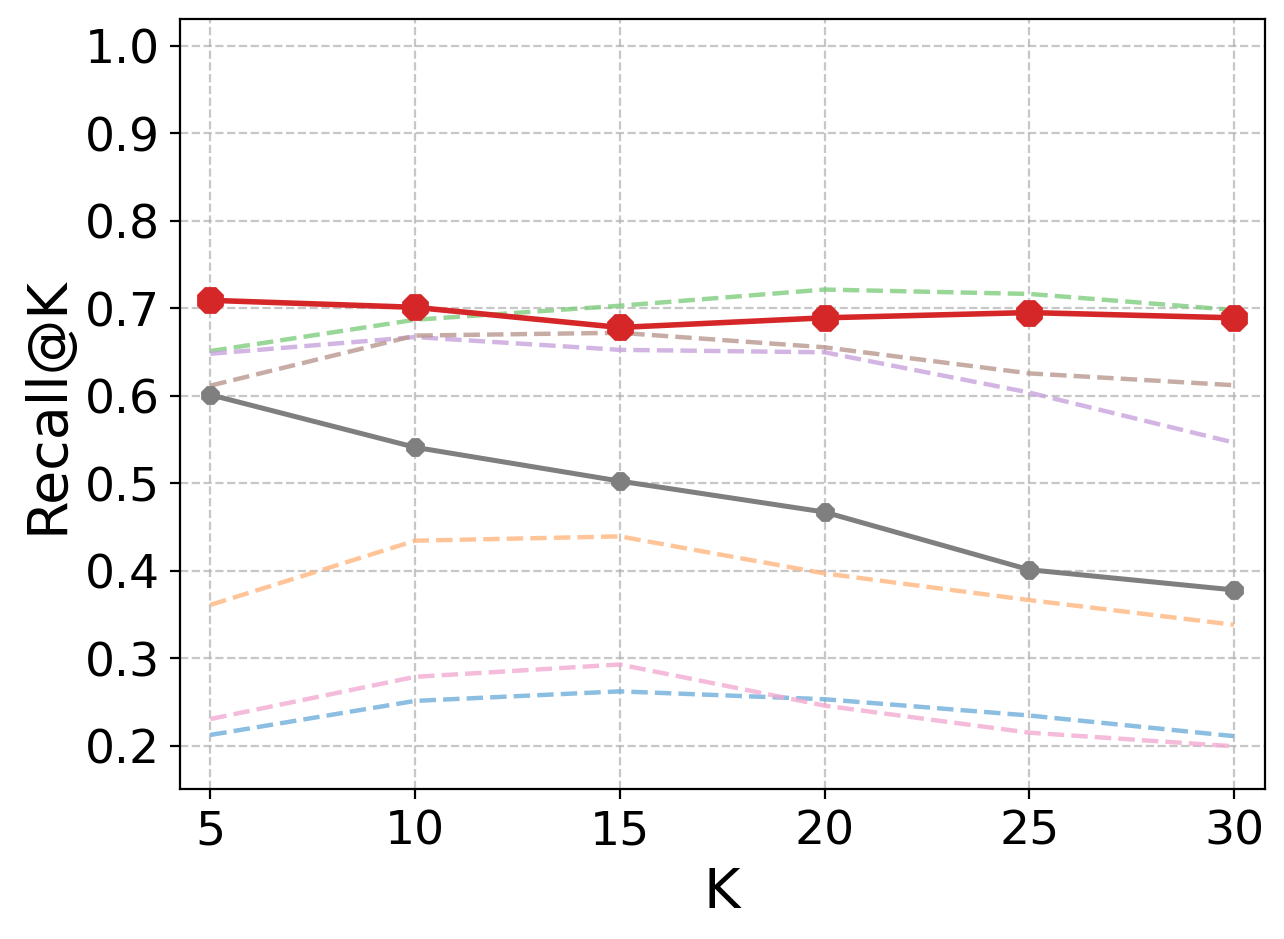}
        \label{fig:om_cr_r}
    \end{subfigure}
    \caption{\revision{$P@k$ and $R@k$ for each tool and benchmark}}
    \label{fig:precision_and_recalls_s}
\end{figure}

\begin{figure}
    \centering
    \includegraphics[width=\columnwidth]{img/effectiveness/p_r_legend_extended.png}
    
    \textbf{Non-synthetic data lakes} \\
    \vspace{0.2cm}
    
    \begin{subfigure}{0.49\columnwidth}
        \centering
        \caption{$P@k$ on D3L$_\text{BM}$}
        \includegraphics[width=0.97\textwidth]{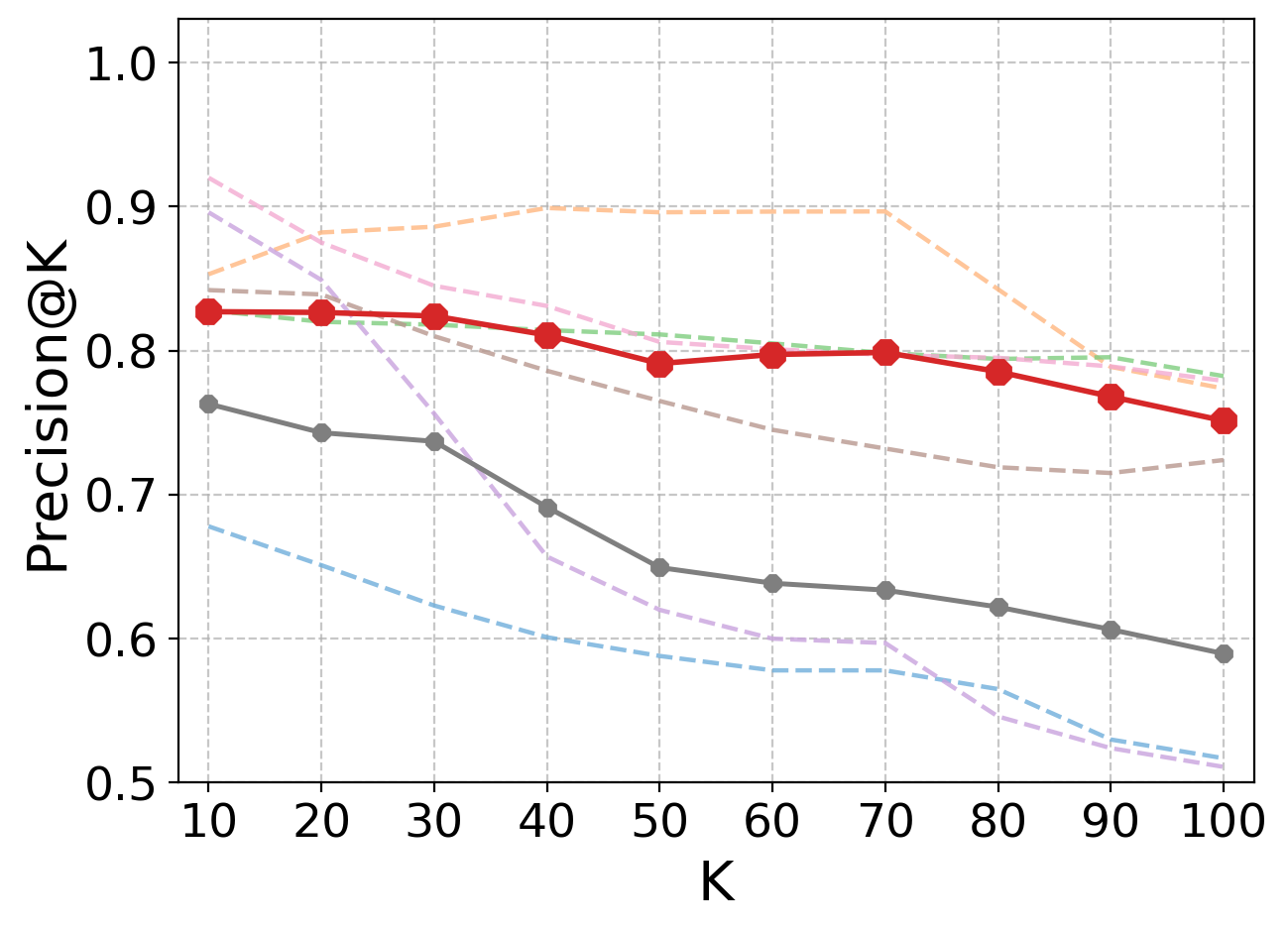}
        \label{fig:d3l_p}
    \end{subfigure}
    \hfill
    \begin{subfigure}{0.49\columnwidth}
        \centering
        \caption{$R@k$ on D3L$_\text{BM}$}
        \includegraphics[width=0.97\textwidth]{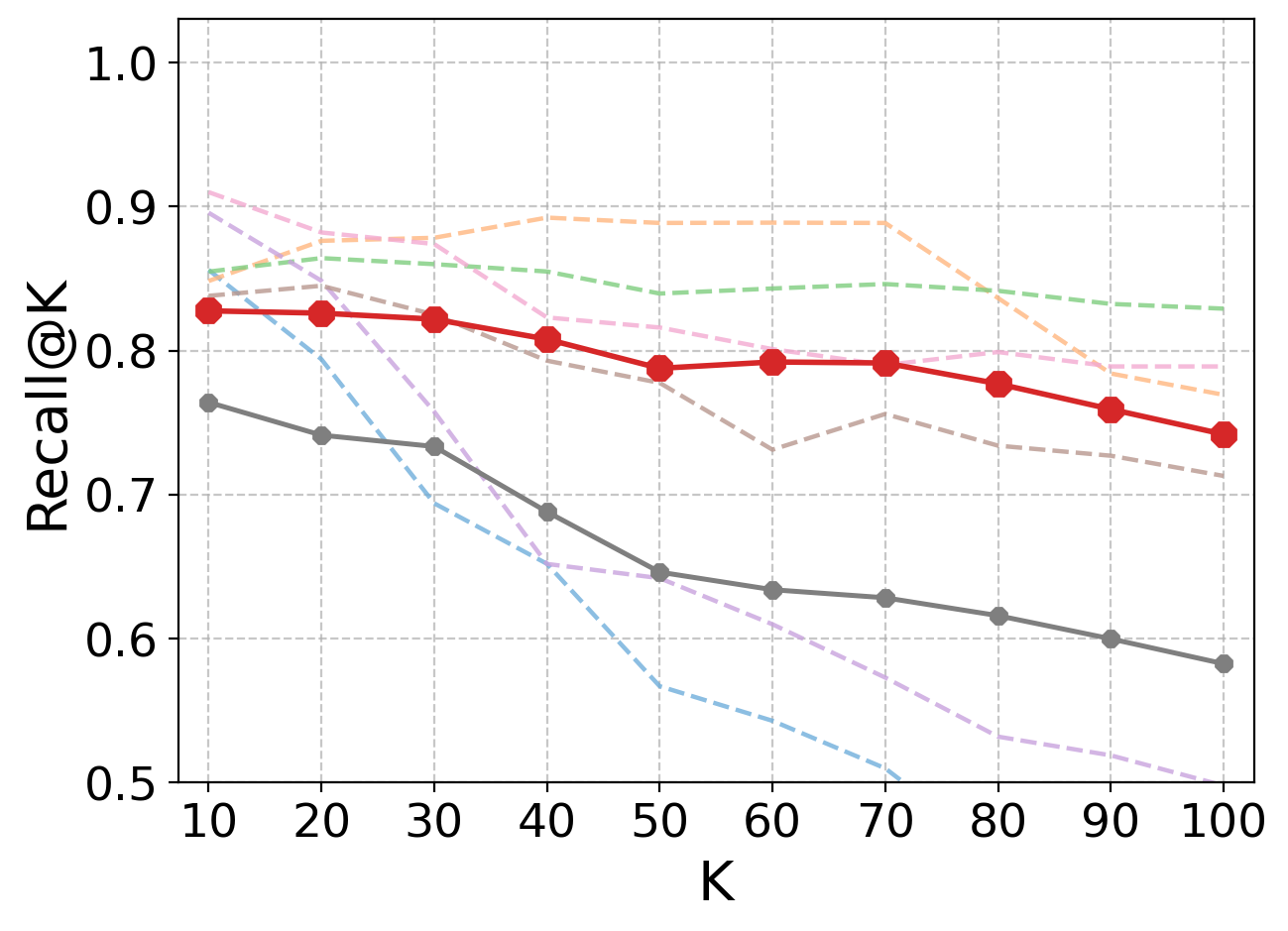}
        \label{fig:d3l_r}
    \end{subfigure}

    \begin{subfigure}{0.49\columnwidth}
        \centering
        \caption{$P@k$ on \sys$_{\text{BM}}$}
        \includegraphics[width=0.97\textwidth]{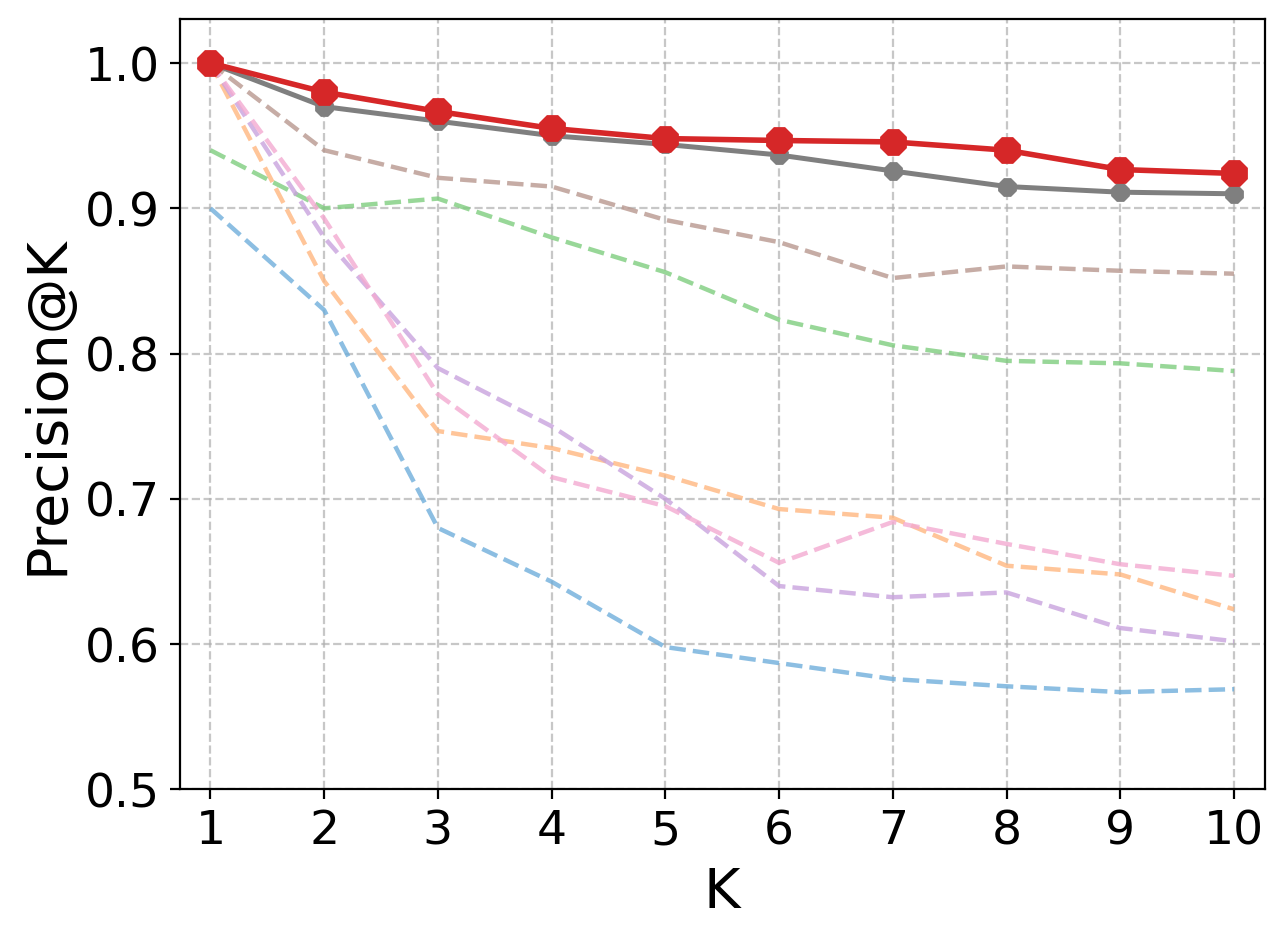}
        \label{fig:nextia_p}
    \end{subfigure}
    \hfill
    \begin{subfigure}{0.49\columnwidth}
        \centering
        \caption{$R@k$ on \sys$_{\text{BM}}$}
        \includegraphics[width=0.97\textwidth]{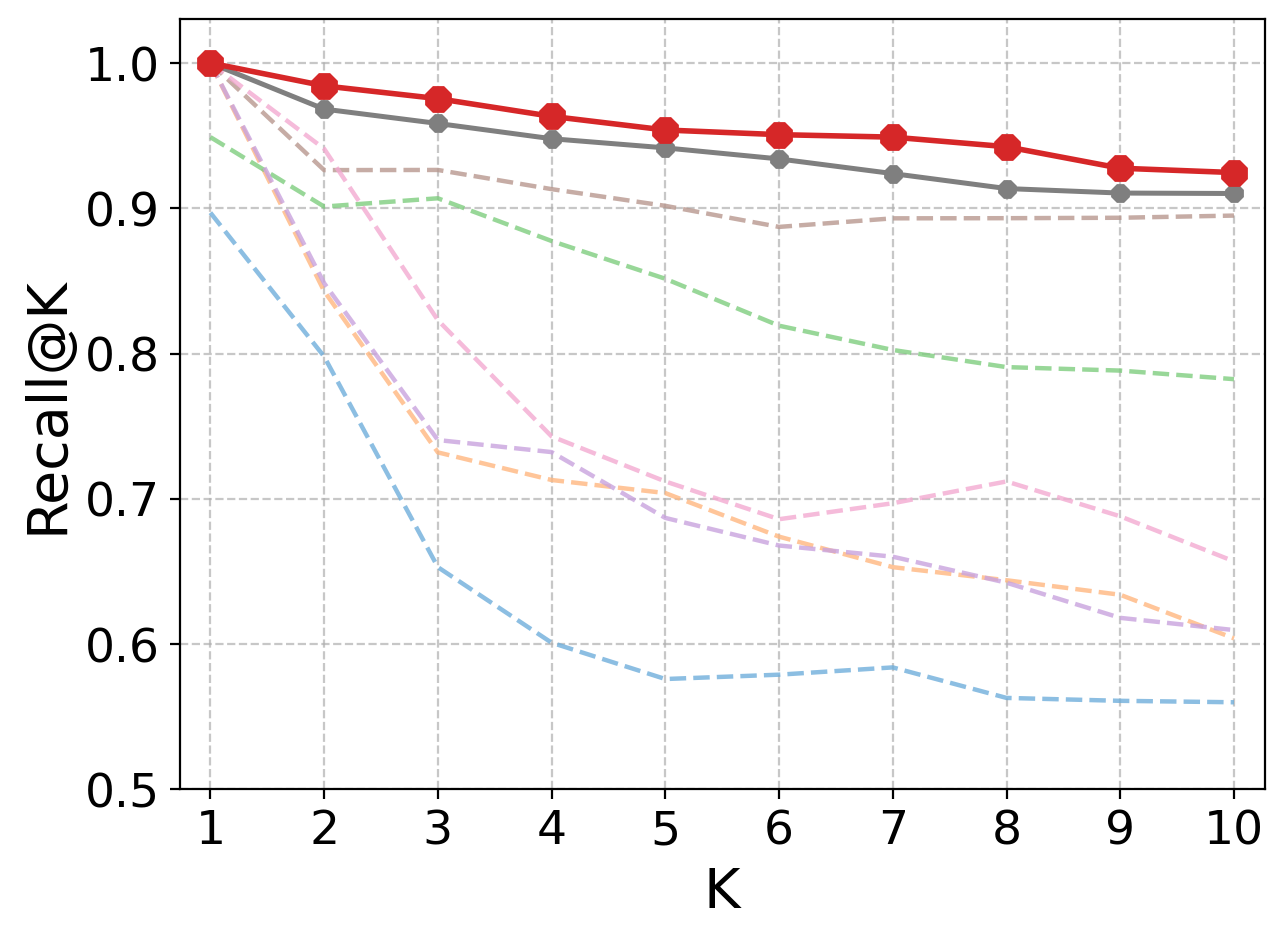}
        \label{fig:nextia_r}
    \end{subfigure}
    \caption{\revision{$P@k$ and $R@k$ for each tool and benchmark}}
    \label{fig:precision_and_recalls_ns}
\end{figure}

\noindent Here, we evaluate the effectiveness of \sys~and its alternatives on all benchmarks except SANTOS Large, since this benchmark lacks ground truth and no state-of-the-art paper reports effectiveness on it. 
Figures \ref{fig:precision_and_recalls_s} and \ref{fig:precision_and_recalls_ns} depict the effectiveness, measured through $P@k$ and $R@k$, of the introduced systems for join detection tasks in the described benchmarks and separated by synthetic and non-synthetic variants. Note that we also included another baseline, $MJ$, obtained from fitting the prediction model described in Section \ref{sec:join_quality_metric} to, rather than the combination of $\mathcal{J}$ and $K$, only $\mathcal{J}$, providing an ablation analysis on the contribution of $K$.

\revision{The main conclusion of Figures \ref{fig:precision_and_recalls_s} and \ref{fig:precision_and_recalls_ns} is that no single system consistently outperforms all others across all benchmarks. Although generalized trends can be observed, such as Starmie and, specially, WarpGate consistently exhibiting a better performance w.r.t. to other approaches, no method dominates in every scenario,} and overall performance varies significantly across benchmarks. The observable variability in the results underscores the inherent difficulty of executing join discovery in data lakes, which further encourages the development of a system that aims at maximizing efficiency whilst maintaining high effectiveness. In both synthetic and non-synthetic data lakes, the performance of \sys~is, at least, comparable to that of other tools, with the only substantial decline happening in the TUS Small benchmark for $20 \leq k \leq 40$. These findings indicate that \sys~enhances efficiency while also preserving, and in some cases exceeding, the join discovery capabilities of semantic approaches.
 
The considerable effort required for manually labeling joins has led the data discovery community to rely heavily on synthetic benchmarks. However, real-world join discovery tasks operate on data lakes whose content is not curated to yield perfectly matching joins. Among the two available non-synthetic benchmarks, \sys~is trained on one (\sys$_{\text{BM}}$) and, as expected, achieves strong performance. Nonetheless, \sys~also performs well on the other non-synthetic benchmark (D3L${_\text{BM}}$), demonstrating its ability to generalize to unseen, real-world contexts. \revision{On the other hand, the OM${_\text{CG}}$ and OM${_\text{CR}}$ benchmarks represent particularly challenging scenarios due to their noise-based data augmentation, but \sys~manages to achieve, respectively, the best and second best average $P@k$ and $R@k$ scores. This highlights the robustness of our approach, which is especially relevant in noisy environments where detecting joinable columns is inherently difficult.}

The experiments conducted demonstrate that \sys~adapts effectively to diverse scenarios while maintaining high performance. It is worth noting that across all benchmarks, \sys~relies on the same predictive model described in Section \ref{sec:predictingQ}. This consistency further illustrates the capability of our profile-based approach to capture and generalize the patterns underlying a wide variety of data lakes, without requiring specific configuration or parameter tuning. Additionally, our approach is fully explainable, as each prediction can be traced back to specific profile features, providing a transparency that is considerably harder to achieve in embedding-based approaches due to the opaque nature of such artifacts.

\stitle{Ablation study} \revision{Figures \ref{fig:precision_and_recalls_s} and \ref{fig:precision_and_recalls_ns} consistently demonstrate that $MJ$, a join prediction model using only the multiset Jaccard metric, performs significantly worse than \sys. After the first few values of $k$, we observe sharp declines in $MJ$’s performance, empirically supporting the hypothesis introduced in Section~\ref{sec:preliminaries}. Incorporating the cardinality proportion substantially mitigates the shortcomings of the multiset Jaccard metric alone, reinforcing its role as a complementary signal that provides lightweight semantic context to an otherwise purely syntactic similarity measure.}


\subsection{Statistical Validation}

\revision{
To ensure the statistical rigor of our experimental findings, we conducted significance testing following the Null Hypothesis Significance Testing (NHST) framework. 
}

\stitle{Efficiency analysis} \revision{Statistical analysis of preprocessing times was not feasible due to the prohibitively high computational costs of repeatedly executing the preparation stage for competing systems, which can take hours or even days per benchmark. However, the substantial differences reported in Table IV (often spanning orders of magnitude) provide clear evidence of \sys's efficiency advantages. For query execution time, we performed $N = 30$ independent runs for each system-benchmark pair and applied one-way ANOVA tests, which rejected the null hypothesis of equal means across all benchmarks ($p < 0.0001$). Tukey's HSD post-hoc tests demonstrated that differences between \sys~and SANTOS/D3L are highly significant ($p < 0.0001$), while differences with embedding-based systems (DeepJoin, KGLiDS, WarpGate) were not statistically significant. All query times in Table IV are reported with 95\% confidence intervals.}

\stitle{Effectiveness analysis} \revision{For effectiveness metrics ($P@k$ and $R@k$), we employed the Wilcoxon signed-rank test, a non-parametric alternative to paired t-tests suitable for bounded metrics that may not follow normal distributions. These tests were conducted across query columns for each benchmark. The results confirmed that performance differences between FREYJA and competing systems are statistically significant across nearly all benchmarks (FREYJA$_\text{BM}$, D3L$_\text{BM}$, TUS Small, TUS Large, OM${_\text{CG}}$, OM${_\text{CR}}$). Notable exceptions where differences were not significant include initial ranking positions ($k = 1, 2$) in SANTOS Small when compared against Starmie, WarpGate, KGLiDS, and DeepJoin, and $k = 1$ in TUS Large against DeepJoin (Precision) and Starmie (Recall). These results validate that the variation in effectiveness scores reflects genuine differences in system performance rather than random execution factors.}

\stitle{Multiple comparisons control} \revision{Given that our evaluation involved 804 statistical tests across multiple benchmarks, metrics, and system comparisons, we addressed the inflated risk of Type I errors (false positives) by applying the Benjamini-Hochberg (B-H) procedure to control the False Discovery Rate (FDR). With an FDR level of $Q = 0.05$, the resulting significance cutoff was determined to be $p \leq 0.0455$. After B-H correction, 765 tests (95.15\% of all tests) remained statistically significant, confirming that the expected proportion of false discoveries among our significant results is controlled at or below 5\%. This validation strengthens the robustness of our experimental conclusions and demonstrates that \sys's performance advantages are not artifacts of random variation but reflect genuine improvements over state-of-the-art approaches.}

%% file: 6_conclusions.tex
\section{Conclusions and future work}\label{sec:conclusions}

\noindent We presented \sys, a novel approach for join discovery in heterogeneous data lakes. \sys~reduces computational costs by orders of magnitude and eliminates the need for high-end hardware, while maintaining state-of-the-art effectiveness on synthetic benchmarks and outperforming alternatives on challenging, non-synthetic data. Our approach introduces two key innovations. First, a novel join quality metric that combines multiset Jaccard and cardinality proportion to measure semantic relatedness beyond simple set-overlap. Second, a general and lightweight model that predicts this metric from data profiles (succinct data characterizations) guaranteeing efficiency and linear scaling. The model requires no fine-tuning and generalizes well to diverse scenarios.


\revision{While \sys~demonstrates strong performance, we acknowledge an inherent limitation of our profile-based approach. By summarizing full column content into data profiles, some nuanced, context-dependent semantic information is inevitably lost. It is plausible that for ambiguous joins dependent on subtle semantic relationships not captured by statistical and structural properties, sophisticated TRL models could have an advantage. For example, lists of generic identifiers could refer to companies, clients or products, and without further context these might be conflated. However, our extensive empirical results suggest that for the broad and practical task of join discovery in heterogeneous data lakes, data profiles are highly effective and lead to state-of-the-art performance.} 

In future work, our aim is to explore the capabilities of profile-based predictive models. That is, profiles can be conceptualized as lightweight embedding representations that prevent the need to use costly encoder architectures to generate traditional, high-dimensional embeddings. \revision{Orthogonally, we could also explore hybrid approaches that leverage \sys~for efficient, large-scale candidate generation, followed by targeted, resource-intensive TRL analysis on a small set of promising candidates to capture the best of both paradigms.} A promising research line encompasses testing the described methodology for data augmentation, experimenting on whether this approach is also capable of assessing the relevance of augmented columns for an analytical task in a scalable manner. 